\documentclass[pre,twocolumn,showpacs,showkeys,superscriptaddress,preprintnumbers,floatfix,nofootinbib]{revtex4-1}

\usepackage{etex}
\usepackage{ifpdf}
\usepackage{hyperref}
\usepackage{dcolumn}
\usepackage{url}
\usepackage{amsmath}
\usepackage{amscd}
\usepackage{amsfonts}
\usepackage{amssymb}
\usepackage{bm}   
\usepackage{bbm}
\usepackage{verbatim}
\usepackage{stmaryrd}
\usepackage{amsthm}
\usepackage{xcolor}
\usepackage{setspace}
\usepackage{braket}
\usepackage{empheq}

\usepackage{tikz}
\usetikzlibrary{arrows}
\usetikzlibrary{automata}
\usetikzlibrary{decorations.pathreplacing}
\usetikzlibrary{positioning}
\usetikzlibrary{plotmarks}
\usetikzlibrary{calc}
\usetikzlibrary{patterns}
\usetikzlibrary{external}
\usepackage{pgfplots}
\pgfplotsset{compat=newest}

\usepackage{graphicx}

\theoremstyle{plain}    
\theoremstyle{plain}    
\theoremstyle{plain}    
\theoremstyle{plain}    
\theoremstyle{plain}    
\theoremstyle{plain}    
\theoremstyle{plain}    
\theoremstyle{plain}    
\theoremstyle{plain}    
\theoremstyle{plain}    
\theoremstyle{plain}    
\theoremstyle{plain}

\newcommand{\CausalState}   { \mathcal{S} }

\newcommand{\forward}{+}
\newcommand{\reverse}{-}
\newcommand{\forwardreverse}{\pm} 

\newcommand{\FutureCausalState} { {\CausalState}^{\forward} }

\newcommand{\PastCausalState}   { {\CausalState}^{\reverse} }

\newcommand{\lastindex}[2]{
  \edef\tempa{0}
  \edef\tempb{#2}
  \ifx\tempa\tempb
    \edef\tempc{#1}
  \else
    \edef\tempa{0}
    \edef\tempb{#1}
    \ifx\tempa\tempb
      \edef\tempc{#2}
    \else
      \edef\tempc{#1+#2}
    \fi
  \fi
  \tempc
}

\newcommand{\CSjoint}[1][,]{
   \edef\tempa{:}
   \edef\tempb{#1}
   \ifx\tempa\tempb
      \ensuremath{\FutureCausalState\!#1\PastCausalState}
   \else
      \ensuremath{\FutureCausalState#1\PastCausalState}
   \fi
}

\newif\ifpm
\edef\tempa{\forwardreverse}
\edef\tempb{\pm}
\ifx\tempa\tempb
   \pmfalse
\else
   \pmtrue
\fi

\newcommand{\kB}{k_\text{B}}  

\newcommand{\LIverson}{\bigl[ \!\! \bigl[}
\newcommand{\RIverson}{\bigr] \!\! \bigr]}

\newcommand{\invol}{\imath}
\newcommand{\linspan}[1]{\mathcal{B}_{#1}}

\begin{document}


\title{Time Symmetries of Memory\\ Determine Thermodynamic Efficiency}

\author{Alexander B. Boyd}
\email{abboyd@ucdavis.edu}
\affiliation{The Division of Physics, Mathematics and Astronomy, Caltech, 1200 E California Blvd, Pasadena CA 91125}

\author{Paul M. Riechers}
\email{pmriechers@gmail.com}
\affiliation{Complexity Institute and School of Physical and Mathematical Sciences,
Nanyang Technological University, 50 Nanyang Ave, 639798, Singapore}

\author{Gregory W. Wimsatt}
\email{gwwimsatt@ucdavis.edu}
\affiliation{Complexity Sciences Center and Physics Department,
University of California at Davis, One Shields Avenue, Davis, CA 95616}

\author{James P. Crutchfield}
\email{chaos@ucdavis.edu}
\affiliation{Complexity Sciences Center and Physics Department,
University of California at Davis, One Shields Avenue, Davis, CA 95616}

\author{Mile Gu}
\email{ceptryn@gmail.com}
\affiliation{Complexity Institute and School of Physical and Mathematical Sciences,
Nanyang Technological University, 50 Nanyang Ave, 639798, Singapore}

\date{\today}
\bibliographystyle{unsrt}

\begin{abstract}
While Landauer's Principle sets a lower bound for the work required for a computation,
that work is recoverable for efficient computations.  However, practical
physical computers, such as modern digital computers or biochemical systems,
are subject to constraints that make them inefficient---irreversibly
dissipating significant energy.  Recent results show that the dissipation in
such systems is bounded by the \emph{nonreciprocity} of the embedded
computation.  We investigate the consequences of this bound for different types
of memory, showing that different memory devices are better suited for
different computations.  This correspondence comes from the time-reversal
symmetries of the memory, which depend on whether information is stored
positionally or magnetically.  This establishes that the time symmetries of the memory
device play an essential roll in determining energetics. The energetic consequences of time symmetries are particularly pronounced in nearly deterministic computations, where the cost of computing diverges as minus
log of the error rate.  We identify the coefficient of that divergence as the
\emph{dissipation divergence}.  We find that the dissipation divergence may be
zero for a computation when implemented in one type of memory while it's
maximal when implemented with another.  Given flexibility in the type of memory, the dissipation divergence
is bounded below by the average state compression of the computation.  Moreover, we show how to explicitly construct the memory to
achieve this minimal dissipation.  As a result, we find that logically
reversible computations are indeed thermodynamically efficient, but logical
irreversibility comes at a much higher cost than previously anticipated.
\end{abstract}


\date{\today}
\maketitle

\setstretch{1.1}


\section{Introduction}
Modern nonequilibrium thermodynamics has established firm lower bounds on the work required to compute via Landauer's bound \cite{Benn82}.  This is the amount of work that must be invested to compress state space during a computation, and thus preserve the Second Law of thermodynamics.  However, this energy is not irretrievably lost \cite{Saga14a,Maro09a,Parr15a}.  It can be regained through precisely designed quasistatic control.  With sufficient control of the energy landscape underlying memory states, an arbitrary computation can be achieved with zero true dissipation \cite{Boyd17a}.

However, perfect efficiency is elusive, because most physical computations are subject to control restrictions that lead to energy dissipation \cite{Kolc20a}.  These constraints on control include finite computation rate \cite{Jun14a, Siva12a, Zulk14a}, modularity \cite{Boyd17a, Riec18b}, and time symmetry \cite{Riec19a,Wims20a}.  Time symmetry is a particularly limiting control restriction, which applies to ubiquitous computational frameworks, such as biochemical processes or modern computers.  Both lack time-asymmetries in their driving signals \cite{Riec19a}.  

For biochemical computing \cite{Ould17a, Hopf74a, Benn79, Sart15}, chemical reservoirs drive nonequilibrium processes in a thermal system, which corresponds to effectively constant driving. These nonequilibrium steady states are a trivial case of time-symmetric computation. Modern computers are subject to the same limitation, because they are driven by a periodic clock signal, which appears the same under time-reversal \cite{Chan92a, Iyer02a}.  It seems that time-asymmetric driving is the exception rather than the rule.  And due to the thermodynamic cost of time-asymmetry in the Brownian regime \cite{Feng08, Rold10a}, autonomous computing may often require this control constraint for a full accounting.~\footnote{In the Brownian regime, any time-asymmetry in driving would correspond to some external device spending energy just to generate the control sequence.}  

Recent results show that time-symmetrically driven computing that operates on metastable memories has an energetic cost well beyond Landauer's bound \cite{Riec19a, Wims20a}.  This corresponds to the \emph{nonreciprocity} of the computation, which reflects how often transitions are made that have less likely reciprocal transitions \cite{Riec19a}.  The reciprocal of a transition is that same transition viewed under time reversal, flipping both the time ordering of memory states and the sign of time-odd physical variables.  Nonreciprocity implies entropy production and dissipated work because of the thermodynamic cost of irreversibility \cite{Rold10a}.  Reciprocal computations, where each transition is equally likely to its reciprocal, are similar in spirit to logically reversible computations, but a stricter class.  Thus, for most familiar forms of computation, there is a different criterion for thermodynamic efficiency, which implicates not just the computation, but also the time-reversal symmetries of the memory. 

For nearly-deterministic time-symmetrically driven computations, the minimum dissipated work diverges proportionally to the negative log-error.  We identify the proportionality constant as the \emph{dissipation divergence}, which is a measure similar to nonreciprocity.  We reduce the expression for dissipation divergence of a computation to the trace of an expression that involves only the computation operator and time-reversal operator on the memory states.

We find that, along with the computation, the time-reversal symmetries of the memory states strongly affect the minimum dissipation of a time-symmetrically driven computation.  If memory is stored in positional memory states, which are preserved under time-reversal, cycles of period greater than two are extremely costly.  However, in magnetic memory states, which flip under time-reversal, we see that computations that iterate a cycle of longer length can be executed without dissipation, according to this bound.  Conversely, we see computations which are costly for magnetic memory but efficient for positional memory, suggesting that different memory devices are better suited for different tasks.

This begs the question, for a particular computation:  How low can we go?  If we choose the right memory device, can we dissipate zero energy, circumventing the apparent limitations of time-symmetric computing?  We answer this question by proving, given total freedom in time-reversal symmetries of the memory, that the dissipation divergence is bounded below by the nullity of the computation matrix divided by the total number of states.  This is the same as the fraction of states with no antecedent in the computation.  Thus, only invertible computations can be executed without divergent dissipation.  Moreover, we explicitly show how to construct a memory with time-reversal symmetries that satisfy this bound.

Thus, while nonreciprocity replaces Landauer's bound as a measure of energy consumption for most computers, 
we see that it recommends logically reversible (invertible) computations just the same.  
However, for these time-symmetrically controlled computations, the energetic cost of logical irreversibility is irretrievable, and increases without bound as the fidelity of our computation increases.  
Thus, logical reversibility's importance is elevated for computations in biochemical media and modern digital computers.

\section{Computational Dissipation of Time Symmetric Control}

For an erasure implemented in a physical system $\mathcal{S}$ whose environment is at temperature $T$, Landauer used the Second Law of thermodynamics to argue that this operation should require at least heat of $\kB T \ln 2$ to compensate for the decrease in system entropy \cite{Land61a}, where $\kB$ is Boltzmann's constant.  
This tradeoff between system entropy and heat bath entropy suggests that irreversible computations require a net expenditure of energy, while reversible computations don't.  This bound has been generalized by setting a lower bound for the average work invested \cite{Parr15a}
\begin{align}
\langle W \rangle \geq \Delta F^\text{neq},
\end{align}
where the nonequilibrium free energy is the average energy of the system minus the temperature times the nonequilibrium entropy of the system
\begin{align}
F^\text{neq}(t)=\langle E(t) \rangle -T S(t).
\end{align}

While Landauer's bound on the energy requirements of computing is informative, equality is only achieved when there is no entropy production and the computation is thermodynamically efficient.  Often, computing requires control restrictions like finite time \cite{Zulk14a, Siva12a}, which yield inefficiencies.  This inefficiency is quantified by the work beyond the nonequilibrium free energy, known as the dissipated work
\begin{align}
\langle W_\text{diss}\rangle = \langle W \rangle -\Delta F^\text{neq}.
\end{align}
In other words, the average dissipated work $\langle W_\text{diss} \rangle$ is the energy required to compute exceeding Landauer's bound.

As shown in Ref.~\cite{Riec19a}, the minimum work required to execute a computation with time-symmetric control considerably exceeds Landauer's bound.  If the memory states $m \in \mathcal{M}$ are a coarse graining ($m \subset \mathcal{S}$, $\bigcup_{m \in \mathcal{M}} m =\mathcal{S}$, and $m \cap m' =  \varnothing$ if $m \neq m'$) of physical states $s \in \mathcal{S}$ and metastably store information, then it was proven that the minimum dissipated work required to compute is
\begin{align}
\label{eq:DissipatedWork}
\beta \langle W_\text{diss} \rangle^\text{t-sym}_\text{min}&=\Delta H_\mathcal{M}
\\&+\sum_{m, m' \in \mathcal{M}}\langle m|\mu_0\rangle p_{m \rightarrow m'} \ln \frac{p_{m \rightarrow m'}}{p_{\dagger(m') \rightarrow \dagger(m)}}. \nonumber
\end{align}
Here, $\beta=1/\kB T$ captures the energy scale of fluctuations in the thermal environment at temperature $T$, $\Delta H_\mathcal{M}$ is the change in Shannon entropy of the memory measured in nats,  $\langle m|\mu_0\rangle$ is the probability of the memory state $m$ initially, and $p_{m \rightarrow m'}$ is the probability of the outcome memory state $m'$ given the input $m$ to the computation.  The set of memory transition probabilities $\{ p_{m \rightarrow m'} \}_{m, m'}$ fully characterizes the computation.  The second term on the right-hand side of the expression, which dominates, is the state-averaged non-reciprocity.  This term is large for computations in which it is common to make transitions $m$ to $m'$, meaning $p_{m \rightarrow m'}$ is large, while the reciprocal transition unlikely, meaning $p_{\dagger(m') \rightarrow \dagger(m)}$ is small.  Here, $\dagger(m)=\{ \dagger(s)|s \in m\}$ is the conjugate memory state of $m$,
while $\dagger(s)$ is the time-reversal of the microstate $s$ which reverses the direction of momenta and magnetic moments.

\begin{figure}[tbp]
\centering
\includegraphics[width=.8\columnwidth]{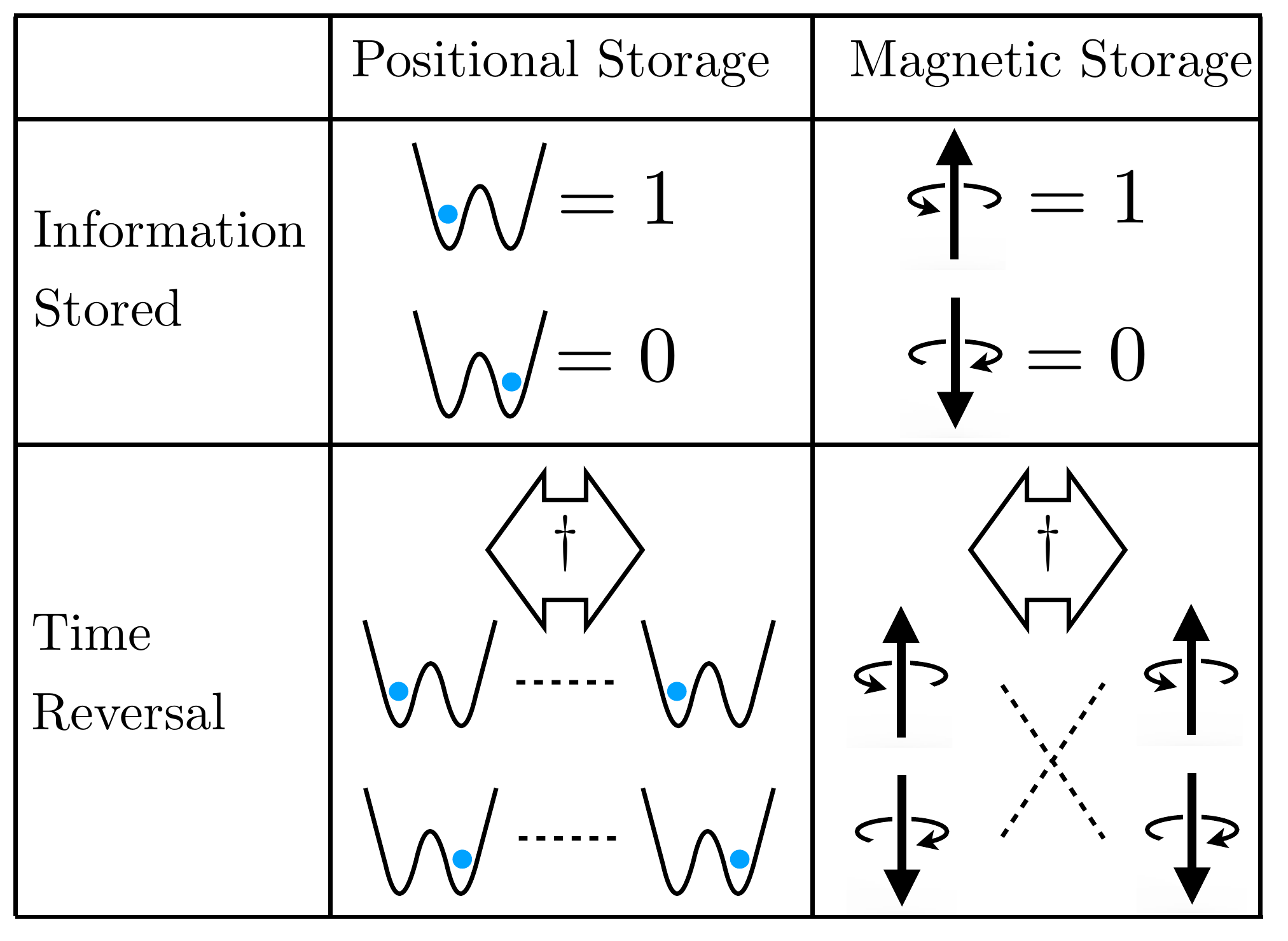}
\caption{TOP LEFT: Positional storage, where a blue particle is in one of two memory states, each given by a positional region: $0$ denoting the left, and $1$ denoting the right.  The energy barrier between them ensures metastability in each by preventing frequent transitions from one to the other.  BOTTOM LEFT: The conjugation operation $\dagger$, which reverses time, leaves these states unchanged, meaning that it is the identity $\hat{\dagger}=\hat{I}$.  TOP RIGHT: Magnetic storage, with up-magnetization corresponding to $0$ and down-magnetization corresponding to $1$. BOTTOM RIGHT: Since the magnetic moment is a time-odd variable, the conjugation operation flips $0$ to $1$ and vice versa.}
\label{fig:MagVSPosConjugate}
\end{figure}

Conjugation comes from the time-reversal of computation, which features in the calculation of entropy production from the detailed fluctuation theorem \cite{Croo99a,Jarz00}.  However, the conjugation can be ignored in many information storing systems, whenever the conjugate of the memory state is itself $m=\dagger(m)$.  This is the case in many systems that are explored in the thermodynamics of control \cite{Jun14a}, where the memory states partition positional variables, as shown in Fig.~\ref{fig:MagVSPosConjugate}, which don't change under time-reversal.

Despite the convenience of the assumption that memory states are preserved under conjugation, and its frequency of use in theoretical and experimental explorations of thermodynamics, practical information storage is often magnetic.  Magnetic fields flip under time reversal, meaning that if we store information with magnetized subregions of a material, then the magnetization flips under time-reversal, as does the stored information.
As shown in Fig.~\ref{fig:MagVSPosConjugate}, if we coarse grain up-magnetized states into a memory state $m$, then conjugation doesn't map $m$ to itself.  Rather, the conjugate memory state $\dagger (m) \neq m$ corresponds to down-magnetized states, which we include as its own memory state: $\dagger(m)\in \mathcal{M}$.
Positional storage and magnetic storage therefore represent two
extremes which demonstrate the importance of the type of memory storage to the
energy required for computation.

\section{Reliable Computation}

In Eq.~(\ref{eq:DissipatedWork}) we see that the same computation,
characterized by the collection of memory-transition probabilities $\{ p_{m \rightarrow m'} \}_{m, m'}$, 
can have very different bounds on the work requirements depending on the time reversal symmetries of $m$ and $m'$.  
This difference is particularly extreme for computations which are \emph{reliable} (nearly deterministic), where there exists a deterministic map $C: \mathcal{M} \rightarrow \mathcal{M}$ such that $m$ maps to $C(m)$ with very low error $\epsilon \ll 1$ so that $p_{m \rightarrow C(m)} \geq 1- \epsilon$.
In Ref.~\cite{Riec19a} it is shown that for such low-error computations, the dissipated work is approximately proportional to the logarithm of the inverse-error: 
\begin{align}
\beta \langle W_\text{diss} \rangle_\text{min}& \approx \Delta H_\mathcal{M} \nonumber
\\ &+ \ln (\epsilon^{-1}) \sum_{m \in \mathcal{M}} \langle m|\mu_0\rangle \LIverson C(\dagger(C(m))) \neq \dagger(m) \RIverson \nonumber
\\ & \equiv \beta \langle W_\text{diss} \rangle^\text{approx}_\text{min}
\label{eq:ApproxWork}
\end{align}
where $\LIverson \cdot \RIverson$ is the Iverson bracket,
which is equal to one if its argument is true and is zero otherwise.  
If we discount the change in memory entropy,
the approximate bound on the dissipated work $ \langle W_\text{diss} \rangle^\text{approx}_\text{min}$ is proportional to the average of this Iverson bracket, which provides the frequency with which the computation makes a transition that is not reciprocated by the time reversal of the computation.  Here, note that we refer to the deterministic map $C$ as the computation, because it approximately characterizes the nearly deterministic computation: $p_{m \rightarrow m'} \approx \delta_{m', C(m)}$.
Ref.~\cite{Wims20a} demonstrates that Eq.~\eqref{eq:ApproxWork} is a good approximation for low error and, moreover, 
validates Eq.~\eqref{eq:DissipatedWork} as a tight bound on the required work for explicit computational models.

The divergent dissipated work in Eq.~(\ref{eq:ApproxWork}) represents a general error-dissipation tradeoff that has been recognized in a variety of biochemically inspired error-correction systems \cite{Benn79, Ould17a, Sart15}.  However, beyond the error rate $\epsilon$, the dissipation depends sensitively on both the target computation $C$ and time reversal operator $\dagger$.  The dependence of the computation on $C$ and $\dagger$ is captured in averaged terms in the Iverson bracket, which specifies the rate at which the dissipation diverges.  We simplify the expression for dissipated work by noting that time reversal enacted twice is the identity $\dagger(\dagger(m))=m$, and that $\LIverson a \neq b \RIverson =1 - \LIverson a=b \RIverson$:
\begin{align*}
\beta \langle W_\text{diss} \rangle^\text{approx}_\text{min}=& 
\\\Delta H_\mathcal{M} + \ln (\epsilon^{-1})  &\sum_{m \in \mathcal{M}} \! \langle m|\mu_0\rangle \bigl( 1- \LIverson \dagger(C(\dagger(C(m)))) = m \RIverson \bigr).
\end{align*}
In what follows, we examine how the dissipation depends on the interaction between the target computation $C$ and the time reversal symmetries of the memory $\dagger$.

It is useful to reframe this result via objects of linear algebra.
To accomplish this, let the memory states $\{ \ket{m} \}_{m \in \mathcal{M}}$ form an orthonormal basis.  If we define the linear operators for the forward computation
\begin{align}
\hat{C}=\sum_{m} \ket{C(m)}\bra{m},
\end{align}
the conjugate computation
\begin{align}
\hat{R} & = \hat{\dagger} \hat{C} \hat{\dagger}
\\ & =\sum_{m} \ket{\dagger(C(\dagger(m)))}\bra{m},
\end{align}
where $\hat{\dagger}$ is a linear operator for time-reversal that maps memory states to their conjugates
\begin{align}
\hat{\dagger}=\sum_{m} \ket{\dagger(m)}\bra{m},
\end{align}
and the initial mixed state
\begin{align}
\hat{\rho}_0=\sum_{m}\ket{m} \langle \mu_0|m\rangle \bra{m},
\end{align}
then the minimum asymptotic work production can be expressed as the average distance between the forward then conjugate computation $\hat{R}\hat{C}$ and identity $\hat{I}$
\begin{align}
\beta \langle W_\text{diss} \rangle^\text{approx}_\text{min} & =\Delta H_\mathcal{M}+ \ln (\epsilon^{-1}) \text{Tr}[\hat{\rho}_0(\hat{I}-(\hat{\dagger}\hat{C})^2)]
\\ & = \Delta H_\mathcal{M}+\ln (\epsilon^{-1}) \text{Tr}[\hat{\rho}_0(\hat{I}-\hat{R}\hat{C})] \nonumber
\\ & = \Delta H_\mathcal{M}+ \ln (\epsilon^{-1}) \langle \hat{I}-\hat{R}\hat{C} \rangle_{\hat{\rho}_0}. \nonumber
\end{align}
This allows us to easily calculate how work bounds for a deterministic computation $\hat{C}$ change due to different types of memory.  The dependence on memory is captured in the the time-reversal symmetry operator $\hat{\dagger}$, and so the conjugate computation $\hat{R}$.  The divergent error-dependent term of the dissipation can be minimized to zero for any input distribution if operating the computation with time-reversal is an involution: $(\hat{\dagger}\hat{C})^2=\hat{I}$.  That is, the computation can then be done thermodynamically reversibly, yielding no entropy production.

As we show explicitly in the following sections, for a particular computation $C$, there are thermodynamically preferred types of memory storage devices, characterized by the time-reversal of that memory $\dagger$.  We say that they are preferred because the desired computation $C$ can be applied with less dissipated work.  

For a particular combination of computation and memory device, we characterize the efficiency with the bound on the asymptotic work required assuming a uniform initial distribution over memory sates $\hat{\rho}_0=\hat{I}/|\mathcal{M}|$:
\begin{align}
\beta \langle W_\text{diss} \rangle^\text{approx}_\text{min}= \Delta H_\mathcal{M}+ \ln (\epsilon^{-1}) \frac{\text{Tr}[\hat{I}-(\hat{\dagger}\hat{C})^2]}{|\mathcal{M}|}.
\end{align}
We choose a uniform initial distribution, because we wish to consider the contribution to dissipation from each memory state equally.  $\frac{\text{Tr}[\hat{I}-(\hat{\dagger}\hat{C})^2]}{|\mathcal{M}|}$ is the fraction of memory transitions $m \rightarrow C(m)$ that are not reciprocated by the conjugate computation ($\dagger(C(\dagger(C(m)))) \neq m$), and so contribute to the divergent dissipation.  This measure of nonreciprocity of a nearly deterministic time-symmetrically driven computation is the most significant factor in determining the minimal energetic cost, because it determines how it must diverge as the error goes to zero.

We identify the asymptotic behavior of the dissipation as the \emph{dissipation divergence} coefficient:
\begin{align}
\mathcal{D}(C,\dagger)& \equiv  \lim_{\epsilon \rightarrow 0}\frac{\beta \langle W_\text{diss}\rangle^\text{approx}_\text{min} }{\ln (\epsilon^{-1})} \nonumber
\\ & =\frac{\text{Tr}[\hat{I}-(\hat{\dagger}\hat{C})^2]}{|\mathcal{M}|},
\end{align}
where the change in Shannon entropy of the memory $ \Delta H_\mathcal{M}$ disappears, because it is bounded by the size of the memory.  Thus to minimize the bound on dissipation, we focus on minimizing the dissipation divergence, as doing so will minimize the divergent terms and allow us to design computations which are potentially more thermodynamically efficient by orders of magnitude.  In what follows, we will use this measure of dissipation to show new bounds on the efficiency of computation, distinct advantages to different memory devices for certain computations, as well as a new equivalence between logical reversibility and thermodynamic efficiency, more extreme than proposed by Landauer \cite{Land61a}.

\section{Positional and Magnetic storage}

For positional storage, where the memory states don't change under conjugation, the conjugate operation is the identity $\hat{\dagger}^\text{pos}=\hat{I}$, and thus the conjugate and forward computation are the same $\hat{R}^\text{pos}=\hat{\dagger}^\text{pos}\hat{C}\hat{\dagger}^\text{pos}=\hat{C}$.  This means that the bound on dissipation for positional storage is only dependent on the forward computation
\begin{align*}
\frac{\beta \langle W_\text{diss} \rangle_\text{min}^\text{pos}}{\ln (\epsilon^{-1})} & \approx   \mathcal{D}(C,\dagger^\text{pos})
\\ & = \frac{\text{Tr}[\hat{I}-\hat{C}^2]}{|\mathcal{M}|}.
\end{align*}
The dissipation divergence is only zero when $\hat{C}^2=\hat{I}$.  This restricts the class of thermodynamically efficient low error computations on positional systems to involutions, which, as we will show, is interestingly the same class of operations as conjugation operators $\hat{\dagger}$.  Every involution can be defined by a collection of pairs of swapped memory states ($C(m) = m'$ and $C(m') = m$) and preserved memory states ($C(m)=m$).

By contrast, storing information magnetically such that $\dagger (m) \neq m$ allows efficiency for very different computations.  Magnetic memory allows thermodynamic efficiency when $\hat{R}^\text{mag}\hat{C}=\hat{\dagger}^\text{mag} \hat{C} \hat{\dagger}^\text{mag}\hat{C}=\hat{I}$, because the bound on dissipation for magnetic storage is
\begin{align*}
\frac{\beta \langle W_\text{diss} \rangle_\text{min}^\text{mag}}{\ln (\epsilon^{-1})} & \approx   \mathcal{D}(C,\dagger^\text{mag})
\\ & = \frac{\text{Tr}[\hat{I}-\hat{R}^\text{mag}\hat{C}]}{|\mathcal{M}|}.
\end{align*}
To explicitly compute this, it is important to note that if we have a system of $N$ magnetic dipoles, where the $i$th dipole gives the bit value $b_i=0$ if it is up and $b_i=1$ if it is down, then the memory state can be represented by an $N$-bit number $\#(m)=\sum_{i=0}^{N-1} b_i 2^i$.  Under the conjugation $\dagger^\text{mag}$, each dipole's magnetic moment is individually flipped such that $0 \mapsto 1$ and $1 \mapsto 0$.  Thus, we can apply the mapping $b_i \mapsto 1-b_i$ to each bit $b_i$.   This means that the $N$-bit number $\#(m)$, which represents the memory state, is mapped to
\begin{align*}
  \#(\dagger^\text{mag} (m)) &= \sum_{i=0}^{N-1} \dagger(b_i) 2^i 
  = \sum_{i=0}^{N-1} (1-b_i) 2^i \\
  &= \sum_{i=0}^{N-1} 2^i - \sum_{i=0}^{N-1} b_i 2^i \\
  &= 2^N - 1 - \#(m)
  ~.
\end{align*}
Thus, taking the conjugate corresponds to the operator $\hat{\dagger}^\text{mag}$,
which can be represented in this memory basis as
\begin{align*}
  \hat{\dagger}^\text{mag}=\sum_{n=0}^{2^N-1} \ket{\#^{-1}(2^N-1-n)} \bra{\#^{-1}(n)} 
  ~,
\end{align*}
where $\#^{-1}$ is the inverse function that maps back from the integers $\mathbb{Z}$ to the the set of memory states $\mathcal{M}$.

In what follows, we compare the divergent dissipation term for magnetic and positional storage for a variety of computations, considering all logical operations on 1 bit and 2 bit systems.   We see that, depending on the computation, one form of storage may be much more efficient than the other.

\section{Dissipation of 1-Bit Computations}

\begin{figure*}[tbp]
\centering
\includegraphics[width=1.9\columnwidth]{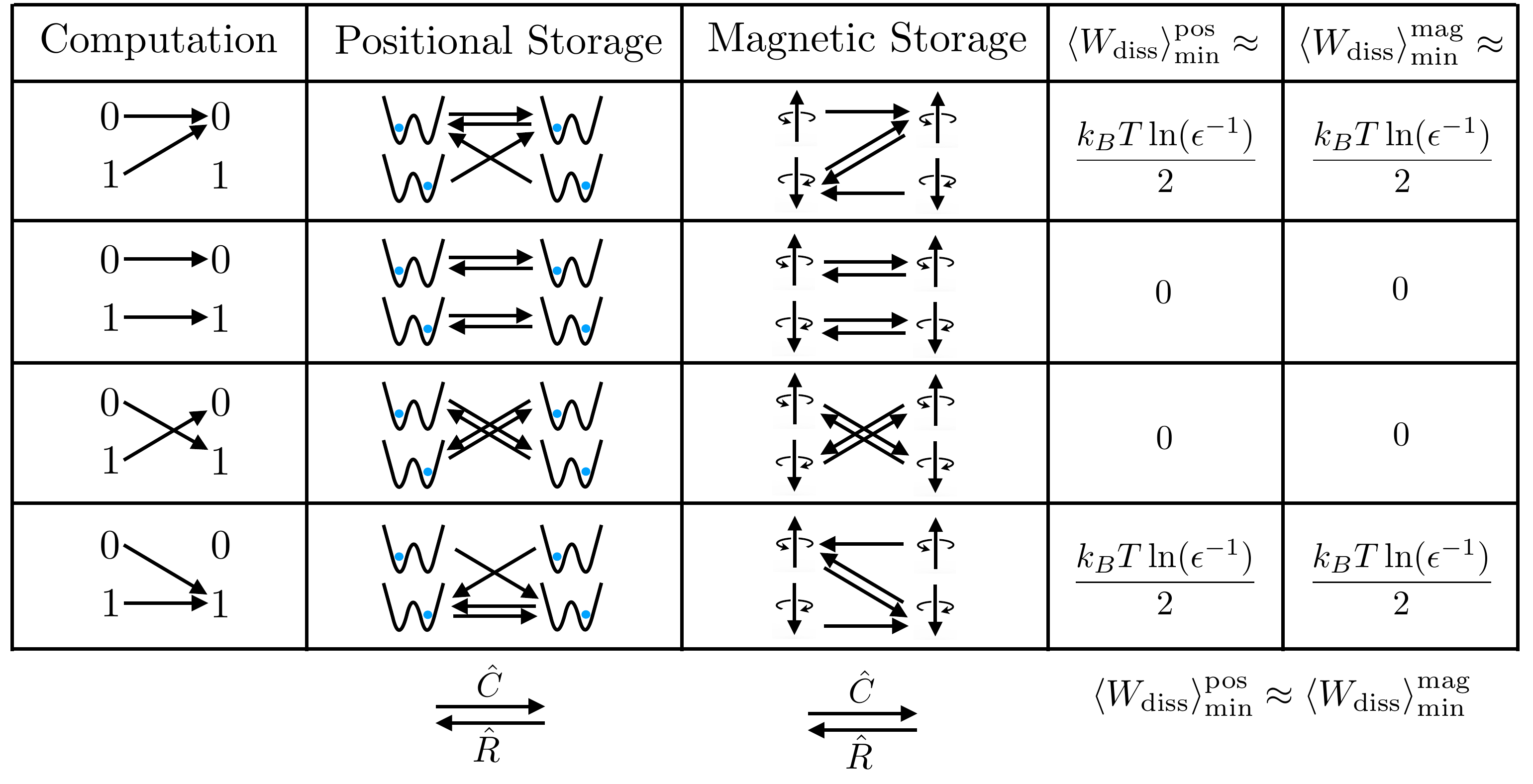}
\caption{From top to bottom, we enumerate all possible deterministic computations on one bit of information in the computation column.
In the positional storage column, we show how each computation $\hat{C}$ maps memory states forward with arrows going from left to right, and how the conjugate computation $\hat{R}$ maps memory states with arrows going from right to left.
Every rightward arrow going from state $m$ to $m'$ that isn't matched by a leftward arrow going from $m'$ to $m$ corresponds to an unreciprocated transition.
The average number of unreciprocated transitions gives the dissipation divergence coefficient, allowing us to approximate the dissipation in the $\langle W_\text{diss} \rangle_\text{min}^\text{pos}$ column.
We match the forward computation and conjugate computation arrows in the same way in the magnetic storage column, and show the approximated dissipation in the $\langle W_\text{diss} \rangle_\text{min}^\text{mag}$ column.
We see that for a single bit, magnetic and positional memory have the same divergent dissipation term for all computations.
}
\label{fig:1BitComputations}
\end{figure*}

Much of the thermodynamics of computation focuses on a single bit of information.  Landauer's bound describes how much energy is required to erase a bit.  However, time-symmetrically controlled positional systems must dissipate much more than the Landauer bound for high-fidelity erasure \cite{Wims20a}.  This prompts the question, does magnetic information storage give an advantage?  Do the different bounds allow for, on average, less work invested and less dissipated?
Fig.~\ref{fig:1BitComputations} shows the results for all 1-bit deterministic computations.

While the identity and swap operations require no dissipation, erasing to $0$ or erasing to $1$ both have dissipation divergence $\mathcal{D}(C,\dagger)=1/2$, requiring $\sim \frac{k_B T}{2} \ln (\epsilon^{-1})$ of dissipated heat in the small error limit, because one of the two transitions is not reciprocated.  This can be seen in both the magnetic and positional column by starting in a state $m \in \{0,1\}$ and following the computation $\hat{C}$ from left to right then following the conjugate computation $\hat{R}$ from right to left.  If this doesn't return the memory to the original state $m$, then starting in this state leads to an energetically costly unreciprocated transition.
For each 1-bit computation, the dissipation divergences for magnetic and positional storage are the same.  This answers our question for 1-bit systems, telling us that there is no advantage in using position or magnetism to execute reliable computations like erasure.
Thus, we must look to 2-bit systems for energetic advantages in different
memory types.

\section{Dissipation of 2-Bit Computations}

\subsection{Advantages of Magnetic and Positional Storage}

\begin{figure*}[tbp]
\centering
\includegraphics[width=1.8\columnwidth]{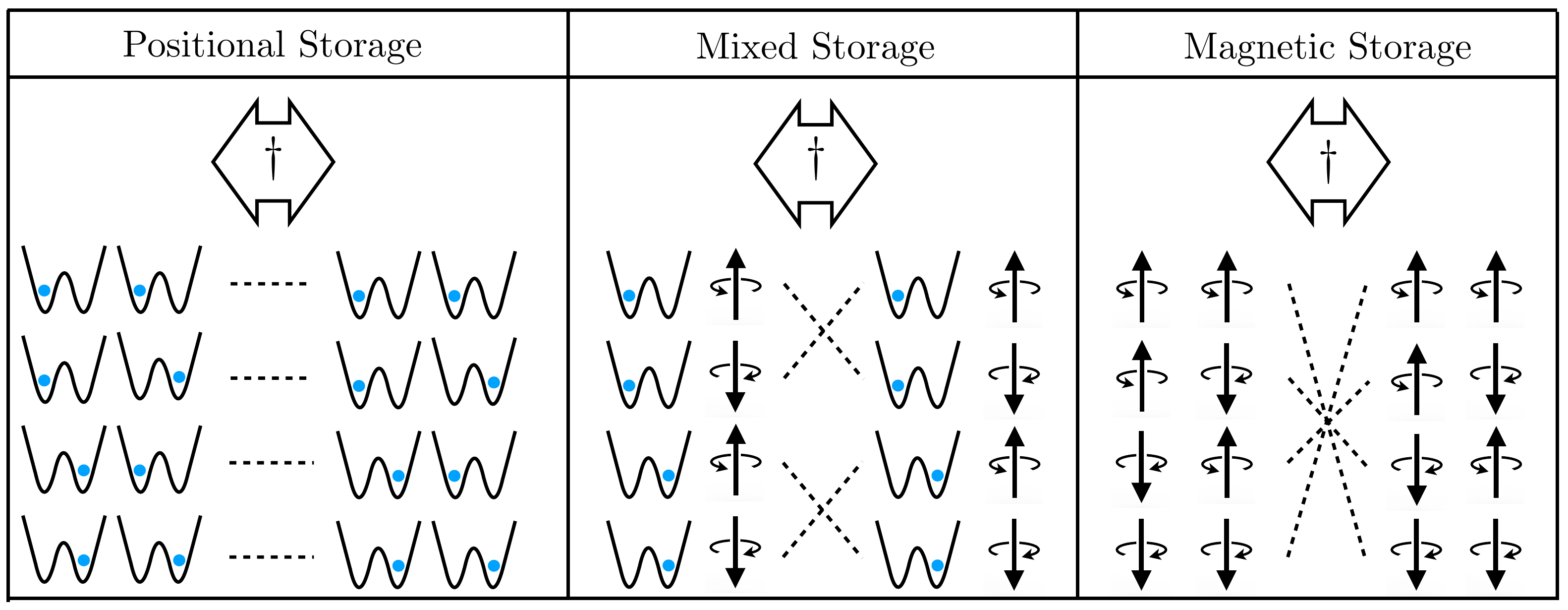}
\caption{Three ways of storing two bits each have a different time-reversal operation (dashed lines from left to right)
LEFT: two positional bistable potentials.  MIDDLE: one positional bistable potential and one magnetic dipole.  RIGHT: two magnetic dipoles. }
\label{fig:ThreeMemoryConjugate}
\end{figure*}

\begin{figure}[tbp]
\centering
\includegraphics[width=\columnwidth]{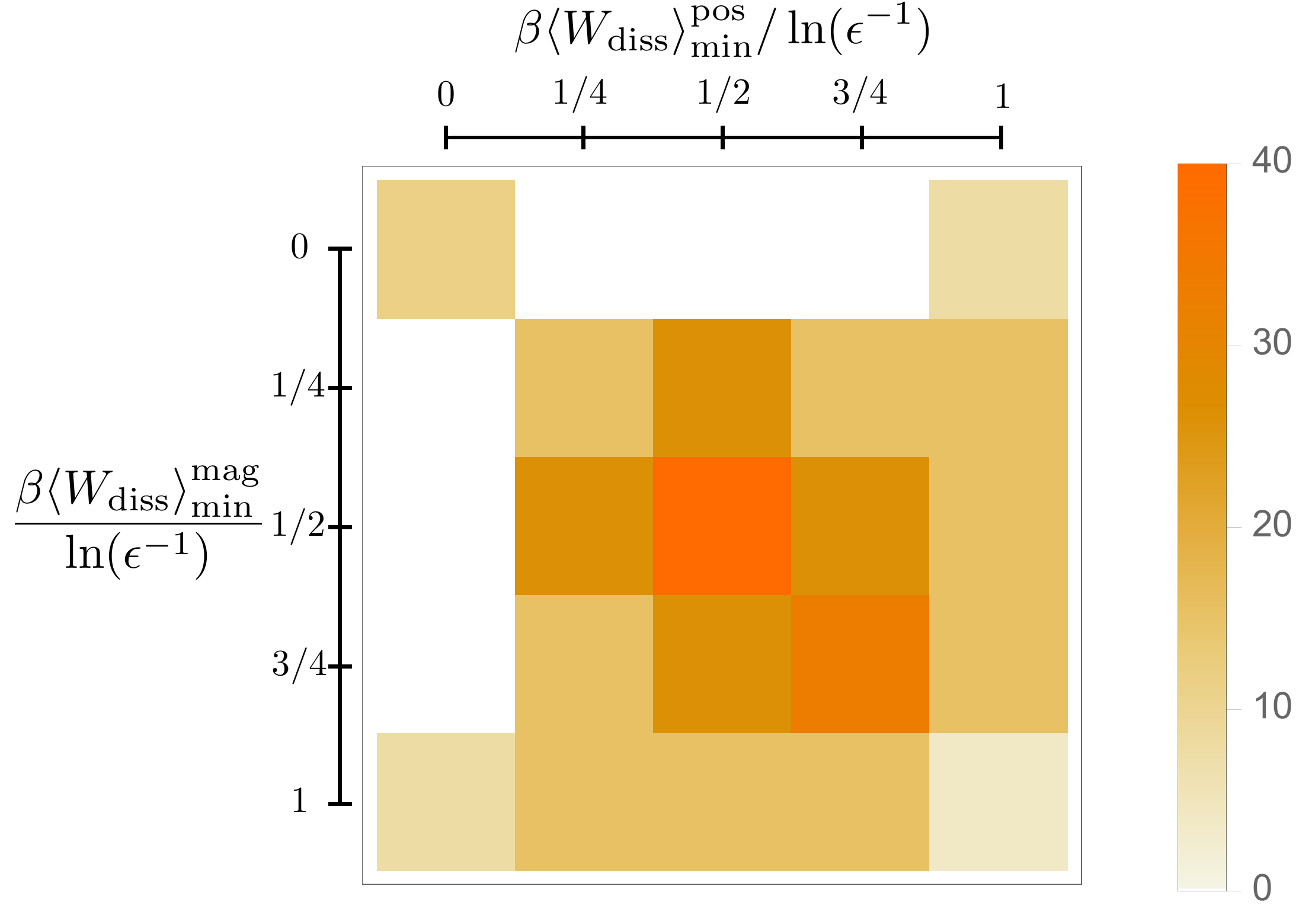}
\caption{The dissipation divergence $D(C,\dagger)=\lim_{\epsilon \rightarrow 0} \frac{\beta \langle W_\text{diss} \rangle_\text{min}}{\ln (\epsilon^{-1})}$ varies depending on whether memory is magnetic or positional (mixed not shown).  Because there are four possible memory states for a computation to act on, the dissipation divergence can be a multiple of $1/4$ within the closed interval $[0,1]$.  In the 5x5 grid above, each square's horizontal location corresponds to dissipation divergence if implemented in positional memory and its vertical location corresponds to dissipation divergence if implemented in magnetic memory.  We plot the number of computations that occupy each element of this grid, revealing that some computations can dissipate less using magnetic memory, while others can dissipate less using positional memory.}
\label{fig:2BitComputations}
\end{figure}

As shown in Fig.~\ref{fig:ThreeMemoryConjugate}, 2-bit memory can be constructed from both magnetic and positional memories, with the time reversal function $\dagger$ given by dashed lines between memory states.  Using these time symmetries, we can evaluate the dissipation of various computations that transform two bits.  However, the suite of deterministic 2-bit computations is larger than can be graphically enumerated in this paper.  Instead, we consider each computation, evaluate the work done for positional and magnetic storage, and plot the counts
for each possible combination of dissipation divergences in Fig.~\ref{fig:2BitComputations}.  Here we note a few key features of the plot:  
\begin{enumerate}
\item There are computations which dissipate more with magnetic storage $\langle W_\text{diss} \rangle_\text{min}^\text{mag} >\langle W_\text{diss} \rangle_\text{min}^\text{pos}$, and computations which dissipate more with positional storage  $\langle W_\text{diss} \rangle_\text{min}^\text{pos} >\langle W_\text{diss} \rangle_\text{min}^\text{mag}$.  Thus, depending on the computation, some forms of information storage can provide energetic advantages over others.  In fact, there are cases, shown in the bottom left and top right corners of Fig.~\ref{fig:2BitComputations}, where one form of storage can execute the computation with the minimal dissipation of zero, but the other form has maximal dissipation divergence.
\item There are computations in the top left corner of Fig.~\ref{fig:2BitComputations}, such as the identity $\hat{C}=\hat{I}$, which can be executed with zero dissipation regardless of how the memory is stored.  Similarly, in the bottom right corner, there are computations which are maximally dissipative regardless of whether the system stores positionally or magnetically.
\item The plot is symmetric with the flip of magnetic and positional storage. Thus, there is no net advantage to magnetic or positional storage when considering all possible operations over two bits.  For every computation which dissipates more with magnetic storage than with positional storage, there is a corresponding computation which dissipates less with magnetic storage,
and vice versa.
\end{enumerate}

\begin{figure*}[tbp]
  \centering
  \includegraphics[width=1.8\columnwidth]{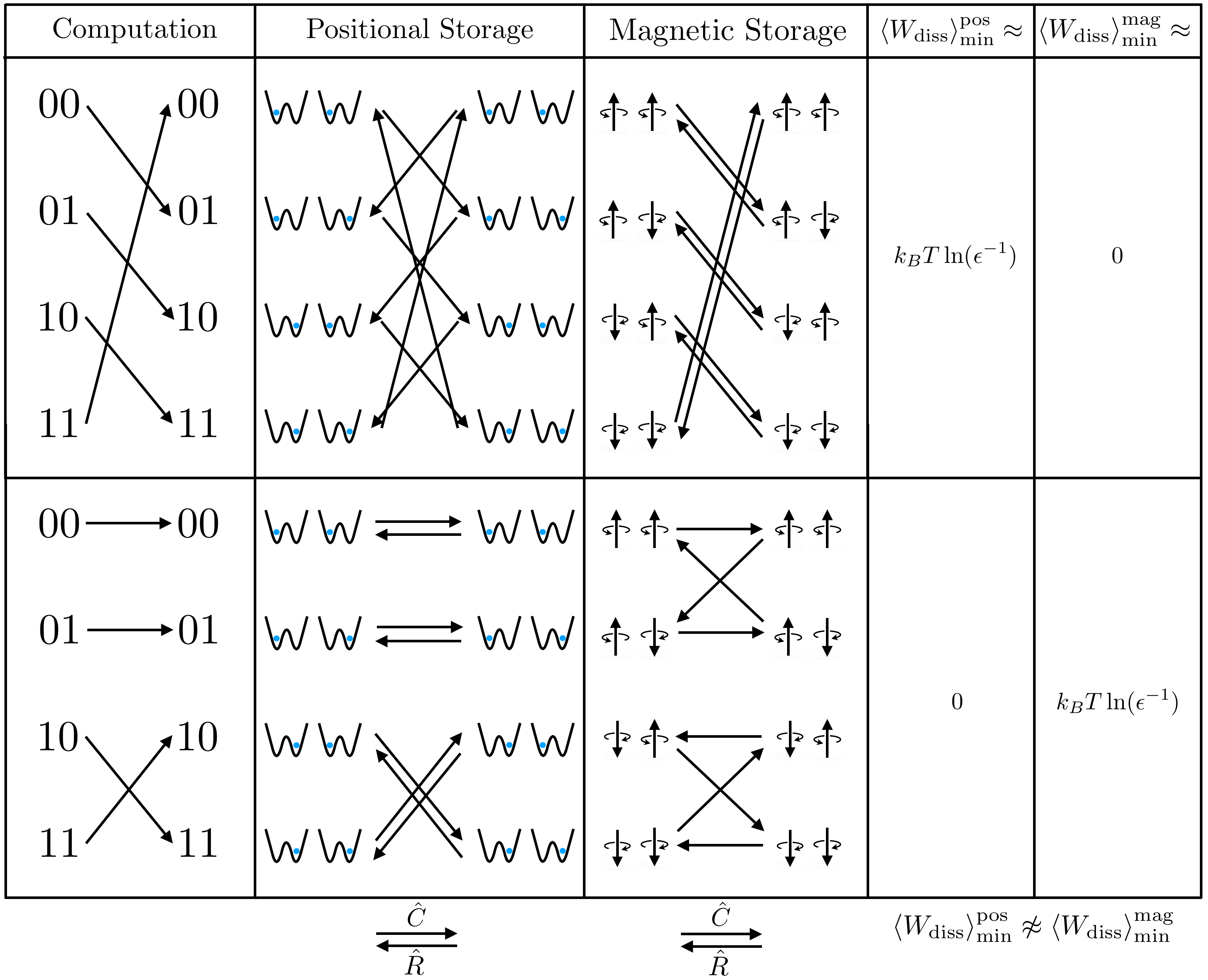}
  \caption{TOP: A 4-cycle computation dissipates maximally when it's implemented with positional storage.  This can be seen from the fact that the forward computation $C(m)=m'$ (arrows from left to right) does not match up with the conjugate computation $R(m')\neq m$ (arrows from right to left).  As a result, $\dagger(C(\dagger(C(m))))\neq m$ for all memory states $m$, and the dissipation divergence is maximized.  By contrast, with magnetic storage, each forward computation $C(m)=m'$ matches up with a conjugate computation $R(m')=m$ so that $\dagger(C(\dagger(C(m))))= m$ for all memory states $m$, and the dissipation divergence is minimized meaning $\langle W_\text{diss} \rangle_\text{min}^\text{pos}\approx k_B T \ln (\epsilon^{-1})$ and $\langle W_\text{diss} \rangle_\text{min}^\text{pos}\approx 0$. BOTTOM: In contrast with the 4-cycle, a computation defined by a single swap of memory states is efficiently implemented with positional storage and dissipates maximally with magnetic storage.}
  \label{fig:2BitSpinVsPosition}
  \end{figure*}

Despite the symmetry of Fig.~\ref{fig:2BitComputations}, there is a functional difference between magnetic storage and positional storage, because the computations which minimize dissipation for one memory type are very different from those that minimize dissipation for the other.  For example, any 4-cycle will be maximally dissipative for positionally stored information.
To see this, first note that none of the memory states return to themselves under a repeated computation $\hat{C}^2$.  As shown in the top half of Fig.~\ref{fig:2BitSpinVsPosition}, this means $\dagger^\text{pos}(C(\dagger^\text{pos}(C(m)))\neq m$ for all $m$.  In turn, the dissipation divergence is maximal: $\mathcal{D}(C, \dagger^\text{pos})=1$.  However, for that same 4-cycle executed with magnetic storage, application of the conjugate computation to the result of the original computation returns the memory to its original state, so that $\hat{R}^\text{mag}\hat{C}=\hat{I}$.
In other words, the computation is reciprocated $\dagger^\text{mag}(C(\dagger^\text{mag}(C(m)))=m$ 
for all $m$.  Thus the dissipation divergence is zero $\mathcal{D}(C,\dagger^\text{mag})=0$, meaning that the dissipation need not diverge as error rate decreases.

Similarly, the bottom half of Fig.~\ref{fig:2BitSpinVsPosition} shows a computation which is very efficient for positional memory but inefficient for magnetic memory.  Every memory element either swaps with another, or remains unchanged.  Thus, when operating the computation twice, we find the identity $\hat{C}^2=\hat{I}$, so that $\dagger^\text{pos}(C(\dagger^\text{pos}(C(m)))= m$ for all $m$.  And so, the dissipation divergence is zero for this computation $\mathcal{D}(C, \dagger^\text{pos})=0$. The reverse computation applied in magnetic storage, on the other hand, does not return a single memory element to itself, so that $\dagger^\text{mag}(C(\dagger^\text{mag}(C(m))) \neq m$ and the dissipation divergence is maximized $\mathcal{D}(C,\dagger^\text{mag})=1$.  Thus, we see that magnetic storage is capable of efficiently implementing computations that would be expensive for positional storage and vice versa.  However, we can look at mixed memory as well to get a broader picture of the space of possibilities.

\begin{figure*}[tbp]
\centering
\includegraphics[width=2\columnwidth]{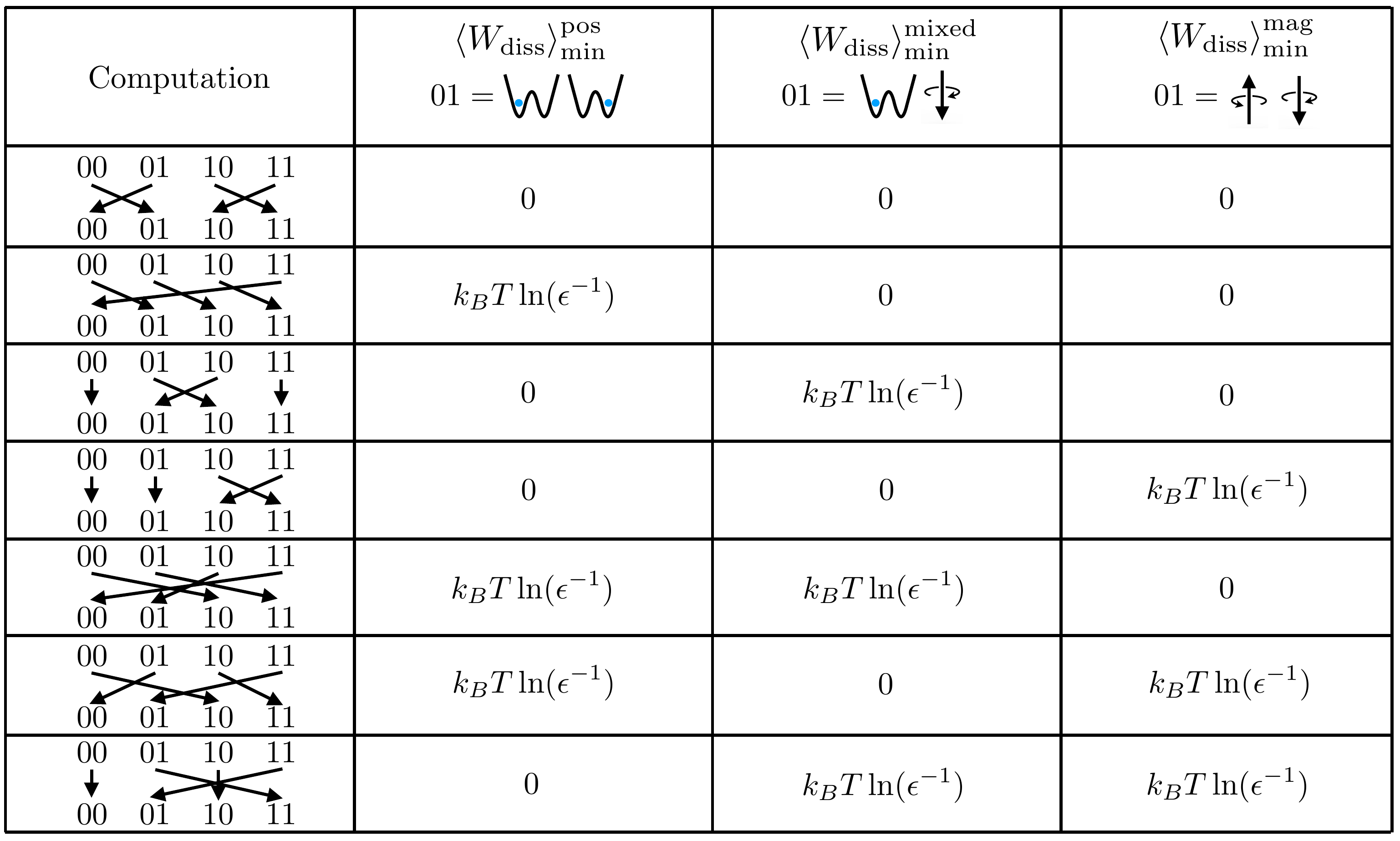}
\caption{We present computations with extreme dissipation divergences over all three forms of memory on two bits in the leftmost column, and show that the dissipation diverges with decreasing error for positional memory (second column from the left), mixed memory (third column from the left), and magnetic memory (last column from the left).  We see all extremal combinations of dissipation except for the case where all three forms of memory dissipate maximally.}
\label{fig:ThreeMemoryComparison}
\end{figure*}

Just as we could have two bits represented with two positional double-wells or two magnetic dipoles, as shown in Fig.~\ref{fig:ThreeMemoryConjugate}, the same figure also shows two bits represented with one positional bistable device and one magnetic dipole.  The different time reversal symmetries shown in the figure affect the thermodynamic bounds.  When we go through all possible computations on four states, we find that all three forms of memory (positional, spin, and mixed) each has its own advantages.  We highlight particular extremal computations in Fig.~\ref{fig:ThreeMemoryComparison}.  Any combination of the three forms of memory can be maximally dissipative while the others are perfectly efficient except for one case: there is no computation for which they all dissipate maximally.  This is suggestive, but doesn't necessarily allow us to determine the best way to implement a computation.

\section{Device Independent Bounds}

Hardware may be designed to be more thermodynamically efficient by tailoring the type of information storage to the intended computation.  
Practically, this would be especially important for designing thermodynamically efficient special-purpose sub-routines,
like the common Fast Fourier Transform (FFT).

The previous section shows that by expanding the class of possible memories, more computations can be executed without divergent dissipation.  There are ways to reduce the cost of computations by changing the physical substrate of computation.  For a given computation $\hat{C}$, the lower bound on dissipation can change radically depending on how the time reversal operator $\hat{\dagger}$ affects the memory states.  This begs the question: is it always possible to compute with zero dissipation if one chooses the right substrate?  As we will see here, the answer is no.  Regardless of the type of memory used to store information and the corresponding time-reversal operator $\hat{\dagger}$, we find that certain types of computations must necessarily dissipate when operated with time-symmetric control.

The baseline dissipation can be found by noting that the trace of a matrix is the sum of its eigenvalues, meaning that we can re-express the dissipation divergence
\begin{align*}
\mathcal{D}(C,\dagger) & \equiv \frac{\text{Tr}[\hat{I}-(\hat{\dagger}\hat{C})^2]}{|\mathcal{M}|}
\\ &=  \frac{|\mathcal{M}|-\sum_{\lambda \in \Lambda_{(\hat{\dagger}\hat{C})^2}} \lambda}{|\mathcal{M}|}
\\ & = \frac{|\mathcal{M}|-\sum_{\lambda \in \Lambda_{\hat{R}\hat{C}}} \lambda}{|\mathcal{M}|},
\end{align*}
where $\Lambda_{\hat{A}}$ denotes the collection of eigenvalues of the matrix $\hat{A}$, each occuring the number of times equal to its algebraic multiplity.  Note that $\hat{\dagger}$, $\hat{C}$, and any composition of them are a subclass of stochastic matrices that are deterministic.  As such, the eigenvalues of $(\hat{\dagger}\hat{C})^2=\hat{R}\hat{C}$ not only have magnitude bounded by $|\lambda | \leq 1$, but they are all either roots of unity, or zero, as shown in App. \ref{sec:Eigenvalues of a Computation}.  This means that the sum of eigenvalues is less than or equal to the number of nonzero eigenvalues:
\begin{align*}
\sum_{\lambda \in \Lambda_{\hat{R}\hat{C}}} \lambda &=\sum_{\lambda \in \Lambda_{\hat{R}\hat{C}},\lambda \neq 0} \lambda+\sum_{\lambda \in \Lambda_{\hat{R}\hat{C}},\lambda=0} \lambda
\\ & \leq \sum_{\lambda \in \Lambda_{\hat{R}\hat{C}},\lambda \neq 0} 1.
\end{align*}
Because the sum of algebraic multiplicities of the eigenvalues is always the
dimension of the underlying space $\sum_{\lambda \in \Lambda_{\hat{R}\hat{C}}} 1=|\mathcal{M}|$, the dissipation divergence is bounded below by the fraction of zero eigenvalues
\begin{align}
\mathcal{D}(C,\dagger)  & = \frac{|\mathcal{M}|-\sum_{\lambda \in \Lambda_{\hat{R}\hat{C}}} \lambda}{|\mathcal{M}|} 
\\ &  \geq \frac{\sum_{\lambda \in \Lambda_{\hat{R}\hat{C}}} 1-\sum_{\lambda \in \Lambda_{\hat{R}\hat{C}},\lambda \neq 0} 1}{|\mathcal{M}|}  \nonumber
\\ &  = \frac{\sum_{\lambda \in \Lambda_{\hat{R}\hat{C}},\lambda = 0} 1}{|\mathcal{M}|} . \nonumber
\end{align}

The nullity of a matrix lower bounds the number of zero eigenvalues, so the nullity describes a lower bound on the dissipation divergence as well:
\begin{align}
\mathcal{D}(C,\dagger)   \geq \frac{\text{nullity}(\hat{R} \hat{C})}{|\mathcal{M}|},
\end{align}
which is the dimension of the input space which maps to the zero vector under the operator $\hat{R}\hat{C}$.  Note that the nullity of the product of matrices is bounded below by the nullity of either of the factors 
\begin{align*}
\text{nullity}(\hat{R}\hat{C})& \geq \text{max}\left[ \text{nullity}(\hat{R}),\text{nullity}(\hat{C})\right]
\\ &\geq \text{nullity}(\hat{C}).
\end{align*}
Finally, we plug into the dissipation divergence and apply the rank--nullity theorem:
\begin{align}
\mathcal{D}(C,\dagger)   & \geq \frac{\text{nullity}( \hat{C})}{|\mathcal{M}|}
\\ & = \frac{|\mathcal{M}|-\text{rank}( \hat{C})}{|\mathcal{M}|} ~. \nonumber
\end{align}

Furthermore, the rank of the computation $\text{rank}(\hat{C})$ is the dimension of the space spanned by the columns of the matrix.  In this case, it is the dimension of the space spanned by the outputs, which is simply equal to the number of possible outgoing memory states for a deterministic mapping
\begin{align}
\text{rank}(\hat{C})&=|C(\mathcal{M})|
\\ & \equiv \bigl| \{ m \in \mathcal{M} | \exists m' \in \mathcal{M} \text{ such that } C(m') = m \} \bigr| ~. \nonumber
\end{align}
Thus, regardless of what form of memory storage is used, the dissipation divergence is at least one minus the fraction of memory space 
in the range of the computation:
\begin{empheq}[box=\fbox]{align}
\lim_{\epsilon \rightarrow \infty} \frac{\beta \langle W_\text{diss}\rangle_\text{min} }{\ln (\epsilon^{-1})}& =\mathcal{D}(C,\dagger) \nonumber
\\ & \geq 1-|C(\mathcal{M})| /|\mathcal{M}|  .
\label{eq:DissipationBound}
\end{empheq}
This bound on the dissipation divergence quantifies how logically irreversible the computation is.  If $C$ is a permutation, mapping onto the entire memory space, then this bound suggests that it could be executed with zero dissipation with the proper memory device.  More precisely, the dissipation divergence bound is the state compression of the memory, reflecting the fraction of memory states that cannot be reached under the computation.  This is conceptually similar to the criterion for irreversibility identified by Landauer's bound, but now with a divergent thermodynamic cost.

However, it remains to be shown whether this device independent bound on dissipation is fundamental. Could it potentially be refined to a tighter bound?  We address this question in the next section. We describe how to design memory devices that, according to the bound set on dissipation by nonreciprocity, can achieve device independent bounds.  Thus, if we allow for total flexibility in how information is stored, Eq.~(\ref{eq:DissipationBound}) represents the strictest possible bound on the cost of computing via the dissipation divergence.

\section{Designing Thermodynamically Efficient Memories}

Here we show how to construct, for any computation $\hat{C}$, a memory device whose dissipation divergence exactly meets the bound of Eq.~\ref{eq:DissipationBound}. To do this, we break the problem down into two parts. Both require that we focus on the device \emph{dependent} term in the equality for the dissipation divergence $ \mathcal{D}(C,\dagger)=\frac{\text{Tr}[\hat{I}-(\hat{\dagger}\hat{C})^2]}{|\mathcal{M}|}$: the time reversal operator $\hat{\dagger}$.  For our purposes, the time-reversal symmetries of the memory characterize all physically relevant aspects of the memory device.  Though, we note that in implementing actual physical computations with real-world constraints, there would naturally be many more factors to consider in achieving the bound.

First, we consider how flexibility in designing memory devices allows for different time-reversal operators.  A time-reversal operator returns a memory state to itself if operated twice: $\dagger(\dagger(m))=m$.  This means that time reversal operations are at least a subclass of involutions.  Fortunately, we also show that every involution can be implemented as a time-reversal operator on some memory device, establishing a formal equivalence between involutions and memory time-symmetries.  We show this by construction, refining a collection of magnetic dipoles into a memory tailored for exactly our needs.  In this way, we present a class of memory devices that gives us maximum flexibility in ensuring an arbitrary time-reversal operator $\hat{\dagger}$ on the memory states.

Second, with the flexibility afforded by our ability to construct a memory device with a time reversal operator given by any desired involution, we investigate how close we can come to the device independent bounds.  In fact, for every computation $\hat{C}$, we show how to construct a time-reversal involution $\hat{\dagger}$ that satisfies the equality $\text{Tr}((\hat{\dagger} \hat{C})^2)=|C(\mathcal{M})|$.  We can therefore meet the bound of Eq.~(\ref{eq:DissipationBound}) for any computation $\hat{C}$.

\subsection{Memories with Flexible Time-Reversal Symmetries}

\label{sec:DesigningInvols}

So far, we have shown a variety of time reversal symmetries by combining bistable systems that are either positional or magnetic.  It turns out that we can construct a memory with \emph{any} involution as its time reversal using \emph{only} magnetic dipoles.  The basic strategy involves two parts:
\begin{enumerate}
\item Construct a sufficiently large memory by combining magnetic dipoles.

\item Redefine the coarse graining by merging memory states until there are the correct number of memory elements that swap, and the correct number that are unchanged by the time reversal.
\end{enumerate}
App.~\ref{app:Construction of Arbitrary Involutions} proves the generality of the method in detail, while we describe the outline of the method here.

To understand the effectiveness of this method, it is important to recognize that \emph{any} involution is composed of two types of operations: self-maps, where states don't change under the involution, and swaps, where two states exchange under the map.  Moreover, a complete set of memory states $\mathcal{M}$ is a coarse graining of the physical state space $\mathcal{S}$ and can be labeled with total flexibility.  Therefore we only need to construct a physical memory with an appropriate number of swapping states and unchanging states and relabel the states according to our desired involution.

\begin{figure}[tbp]
\centering
\includegraphics[width=\columnwidth]{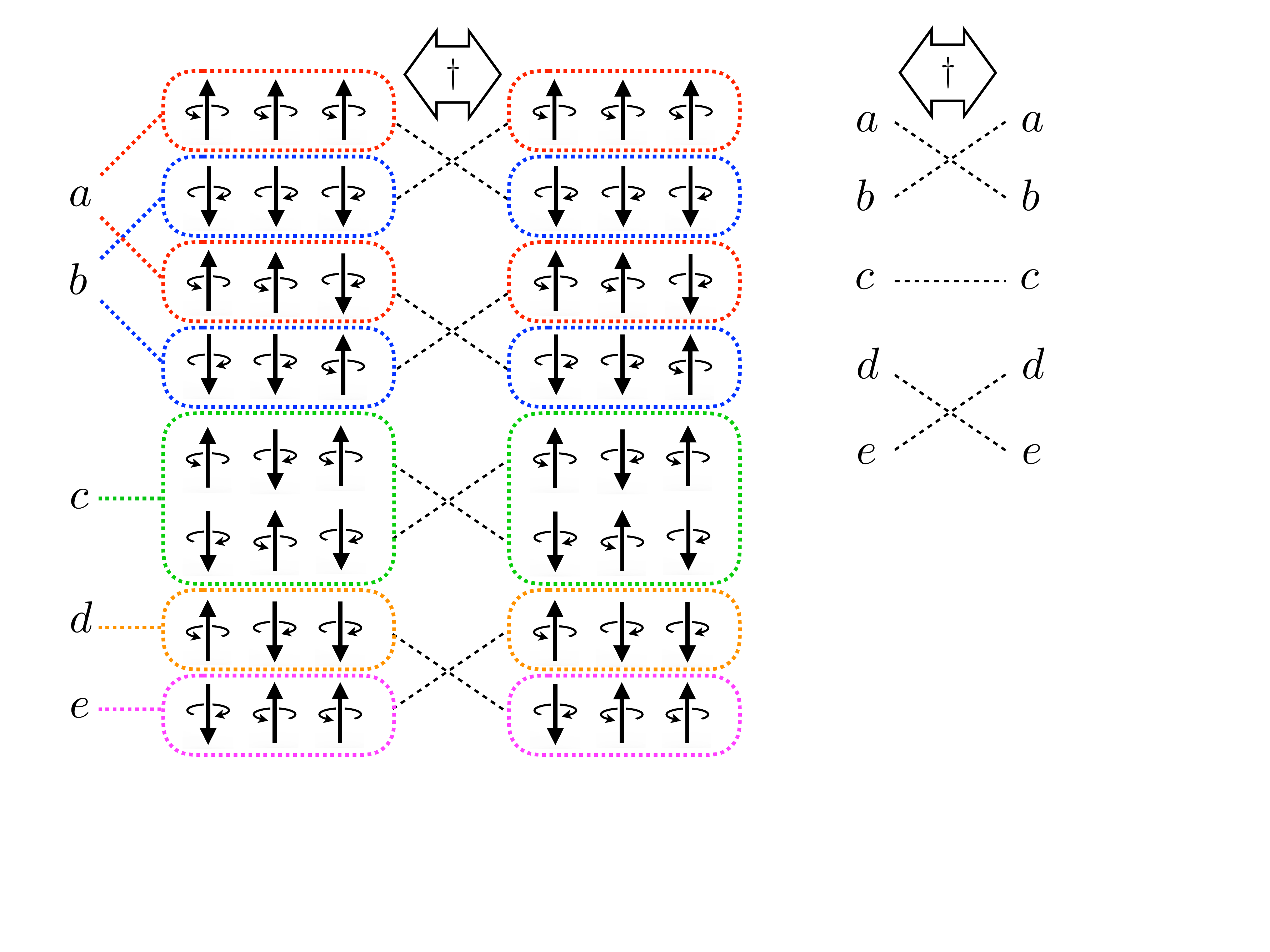}
\caption{By coarse-graining magnetic dipoles, we can implement a memory device with any involution as its time-reversal symmetry among memory states.  In this case, we coarse-grain three magnetic dipoles to yield the memory $\mathcal{M}'= (a,b,c,d,e)$, with the time-reversal $(\dagger(a),\dagger(b),\dagger(c),\dagger(d),\dagger(e))=(b,a,c,e,d)$.  Combinations of spins surrounded by the same color border have been merged into the same memory state. }
\label{fig:MemoryMerging}
\end{figure}

Step 1 in this process, where we construct a memory of $S$ magnetic dipoles, provides $2^S$ memory states as a substrate.  Each combination of up and down orientations of the dipoles corresponds to a particular memory state.  Fig.~\ref{fig:MemoryMerging} shows a memory on the left-hand side that is constructed from three dipoles and thus has $2^3=8$ memory states.  As shown in Fig.~\ref{fig:MemoryMerging}, every memory state $m$ in our dipole memory substrate swaps under time-reversal with a partner state $\dagger(m)$.

We redefine the memory in Step 2 by merging memory states with each other. Merging states $m$ and $m'$ is done by taking the union of their underlying microstates $m \cup m' = \{s| s \in m \text{ or } s \in m'\}$. We can then use the new state $m \cup m' $ to replace $m$ and $m'$ and define a legitimate new set of memory states $\mathcal{M}'= \mathcal{M} \cup \{ m \cup m' \} \setminus \{ m,m'\}$.  $\mathcal{M}'$ is also a coarse graining of the physical microstates $\mathcal{S}$, but with a different time-reversal operator.

Fig. \ref{fig:MemoryMerging} shows how to use this technique to define a physical memory $\mathcal{M}'= (a,b,c,d,e)$, with the time-reversal $(\dagger(a),\dagger(b),\dagger(c),\dagger(d),\dagger(e))=(b,a,c,e,d)$ from the substrate memory that uses three magnetic dipoles to store information $\mathcal{M}=\{ \downarrow , \uparrow \}^3$.  The first task is to create a memory state $c$ which doesn't change under time-reversal.  This can be done by taking two memory states that are each other's time reversal, $\uparrow \downarrow \uparrow$ and $ \downarrow \uparrow \downarrow= \dagger(\uparrow \downarrow \uparrow)$, and merging them into one state $c =\uparrow \downarrow \uparrow \cup  \downarrow \uparrow \downarrow$.  The new memory state $c$ is fixed under time reversal because, when conjugated, any microstate $s$ within $\downarrow \uparrow \downarrow$ maps to a microstate $\dagger(s)$ within $\uparrow \downarrow \uparrow$ which is also within $c$.  The converse is also true for states that map from $\downarrow \uparrow \downarrow$ to $\uparrow \downarrow \uparrow$ under conjugation.  This leaves us with a memory that has one unchanging state $c$ and six memory states that swap under time reversal.  However, our desired memory $\mathcal{M}'$ only has four swapping states, meaning we need to continue merging states of the substrate.

To reduce the number of swapping states from six to four, we again merge memory states.  However, rather than merge two states $m$ and $m'$ that are conjugates of each other $\dagger(m)=m'$, we make sure that $\dagger(m) \neq m'$.  We also coarse grain their conjugates $\dagger(m)$ and $\dagger(m')$ into a new memory state.  Specifically, Fig. \ref{fig:MemoryMerging} shows how we merge $\uparrow \uparrow \uparrow$ and $\uparrow \downarrow \downarrow$ into $a=\uparrow \uparrow \uparrow \cup \uparrow \downarrow \downarrow$, as well as their conjugates into $b=\dagger(\uparrow \uparrow \uparrow) \cup \dagger(\uparrow \downarrow \downarrow)=\downarrow \downarrow \downarrow \cup \downarrow \uparrow \uparrow$.  Any microstate within $a$, contained by either $\uparrow \uparrow \uparrow$ or $\uparrow \downarrow \downarrow$, maps to a microstate within $b$ under time reversal, and visa versa.   This results in two memory states $a$ and $b$ that swap under time reversal $\dagger(a)=b$ taking the place of four states. Finally, we relabel the remaining magnetic states as $d$ and $e$, resulting in the desired memory system $\mathcal{M}'$.

As shown in App.~\ref{app:Construction of Arbitrary Involutions}, this procedure can be generalized. 
App.~\ref{app:Construction of Arbitrary Involutions} demonstrates that it is possible to construct a memory where time-reversal acts as an arbitrary involution on the memory states.
This gives a high degree of flexibility in designing energy-efficient computations.
In the next section, we investigate how to mathematically construct the requisite involution for us to meet the error--dissipation bound of Eq.~\ref{eq:DissipationBound}.

Note that this technique is contingent on a substrate memory that is composed of time-antisymmetric swapping states.  Such states can be merged to produce time-symmetric unchanging memory states, but the converse is not true.  Time-symmetric positional memories cannot be used in this way to create time-antisymmetric memory states.  In the previous sections, we saw no advantage to positional or magnetic memory substrates for general computing.  While we've seen that some computations are better suited for certain types of memory, Fig.~\ref{fig:2BitComputations} shows that, when averaging over all computations, the two types of memory are energetically equivalent.  However, the result described in this section implies that a time-antisymmetric memory that is large enough can be as efficient as any other memory type, because it can be transformed into a memory with the same time-reversal symmetries.  Therefore, we've identified a distinct advantage to storing information with magnetic dipoles over positional forms of memory.

Finally, while strictly positional time-symmetric memories are not flexible enough to access a wide variety of memory symmetries, we need not turn to purely magnetic memories.  As little as a single magnetic dipole $\{ \uparrow ,\downarrow \}$ can be added to a time-symmetric positional memory $ \mathcal{M}$ to make all the memory states states time-antisymmetric.  The resulting memory states of the expanded space $m' \in \mathcal{M'} = \{\uparrow, \downarrow \} \times \mathcal{M}$ each swap with a partner under time reversal.  For instance, for the expanded memory state  $m'= (\uparrow ,m)$, where $m \in \mathcal{M}$ is unchanging, the time reversal yields a different state $\dagger(m')= (\dagger(\uparrow),\dagger(m))=(\downarrow, m)$.  Therefore, the technique described above for constructing any memory symmetry can be applied to such a system.  A small amount of time-asymmetry in memory storage unlocks a vast array of thermodynamic possibilities.

\subsection{Designing Efficient Time-Reversal Symmetries}

As shown in the last section, to make a computation $C$ efficient, we can change the physical substrate in order make time-reversal of the memory states $\dagger$ any involution that we choose.  If the computation is as efficient as possible, then the time-reversal of the memory must achieve the bound shown in  Eq.~(\ref{eq:DissipationBound}), meaning that
\begin{align}
\frac{ \text{Tr}(\hat{I}-(\hat{\dagger} \hat{C})^2)}{|\mathcal{M}|}=\frac{|\mathcal{M}|-|C(\mathcal{M})|}{|\mathcal{M}|} ~. \nonumber
\end{align}
This condition is met if the trace of $(\hat{\dagger}\hat{C})^2$ is the same as the size of the image of the computation:
\begin{align}
\text{Tr}((\hat{\dagger}\hat{C})^2)=|C(\mathcal{M})|.
\end{align}
We will now show how to determine the involution $\dagger$ that pairs with the computation $C$
to achieve this optimal thermodynamic efficiency.

\begin{figure*}[tbp]
  \centering
  \includegraphics[width=1.8\columnwidth]{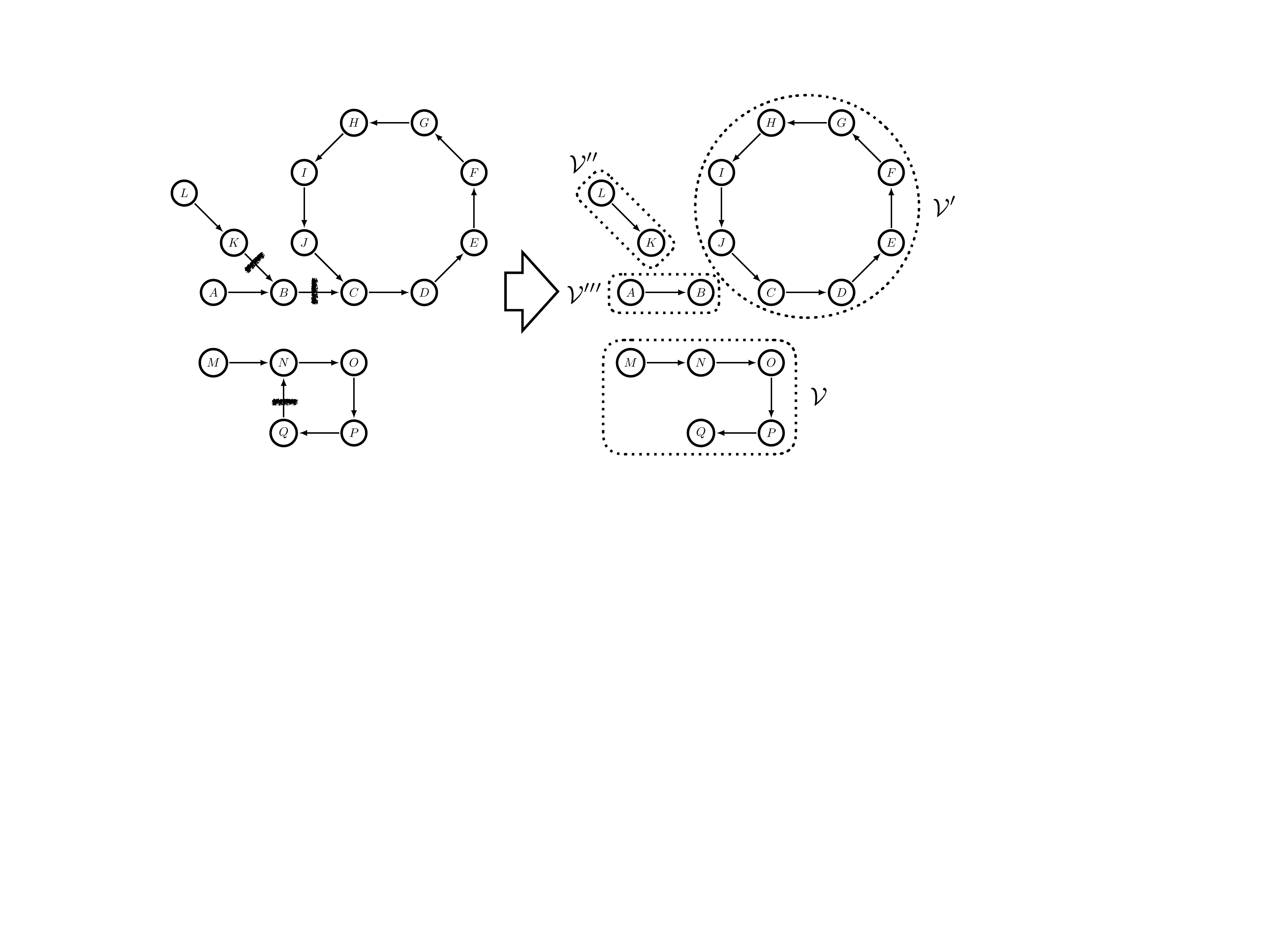}
  \caption{On the left, we have computation $C(m)$ specified over the states $\mathcal{M}=\{A,B,C,D,E,F,G,H,I,J,K,L,M,N,O,P,Q\}$ with a directed graph representing memory transitions.  We prune redundant transitions with the shown scratch marks.  The result is four disconnected components that either have line or loop topology.  In this case, we have three line-topology components $\mathcal{V}=\{M,N,O,P,Q\}$, $\mathcal{V}''=\{L,K\}$, and $\mathcal{V}'''=\{A,B\}$, which means that the computation has nullity$=3$.  The single loop component is $\mathcal{V}'=\{C,D,E,F,G,H,I,J\}$.  }
  \label{fig:GraphPruning}
  \end{figure*}

In designing the time-symmetries $\dagger$ of the memory to efficiently support a particular computation, it is useful to draw out the memory state transitions induced by that computation $m \rightarrow C(m)$ as a directed graph as shown on the left side of Fig.~\ref{fig:GraphPruning}. Specifically, we note that there may be \emph{redundant transitions} $m \rightarrow C(m)$ for which there are other transitions $m' \rightarrow C(m')$ which share the same computational outcome $C(m)=C(m')$.  We call these redundant, because they do not contribute to the size of the image of the computation $|C(\mathcal{M})|$.  Thus, we can prune the graph by cutting redundant transitions while preserving the number of memory states to which the directed graph points.  

As shown in Fig.~\ref{fig:GraphPruning}, we prune 
until every memory state has at most one antecedent state that maps to it under the mapping described by the pruned graph.  This pruning process leaves exactly one edge going to each memory state of the image of the computation $C$.  Thus, the number of cuts made is the total number of initial edges minus the size of the image of the computation $|\mathcal{M}|-|C(\mathcal{M})|$.

What's left after cutting redundant transitions, as shown on the right side of Fig.~\ref{fig:GraphPruning}, will be a new directed graph where each memory state is part of a connected component $\mathcal{V} \subset \mathcal{M}$.  Every state $m$ has at most one destination state $C(m)$ and at most one antecedent state $m'$ such that $C(m')=m$, implying that each component of the pruned graph is either a loop or a line with a start and end state.  These are the only possible topologies for such constraints on connectivity.  The end state of every line component corresponds to a memory state whose redundant transition was pruned.  Thus, the number of line components and the number of end states within the pruned graph are both $|\mathcal{M}|-|C(\mathcal{M})|$, the number of cuts made to prune redundant transitions.

\begin{figure*}[tbp]
\centering
\includegraphics[width=1.8\columnwidth]{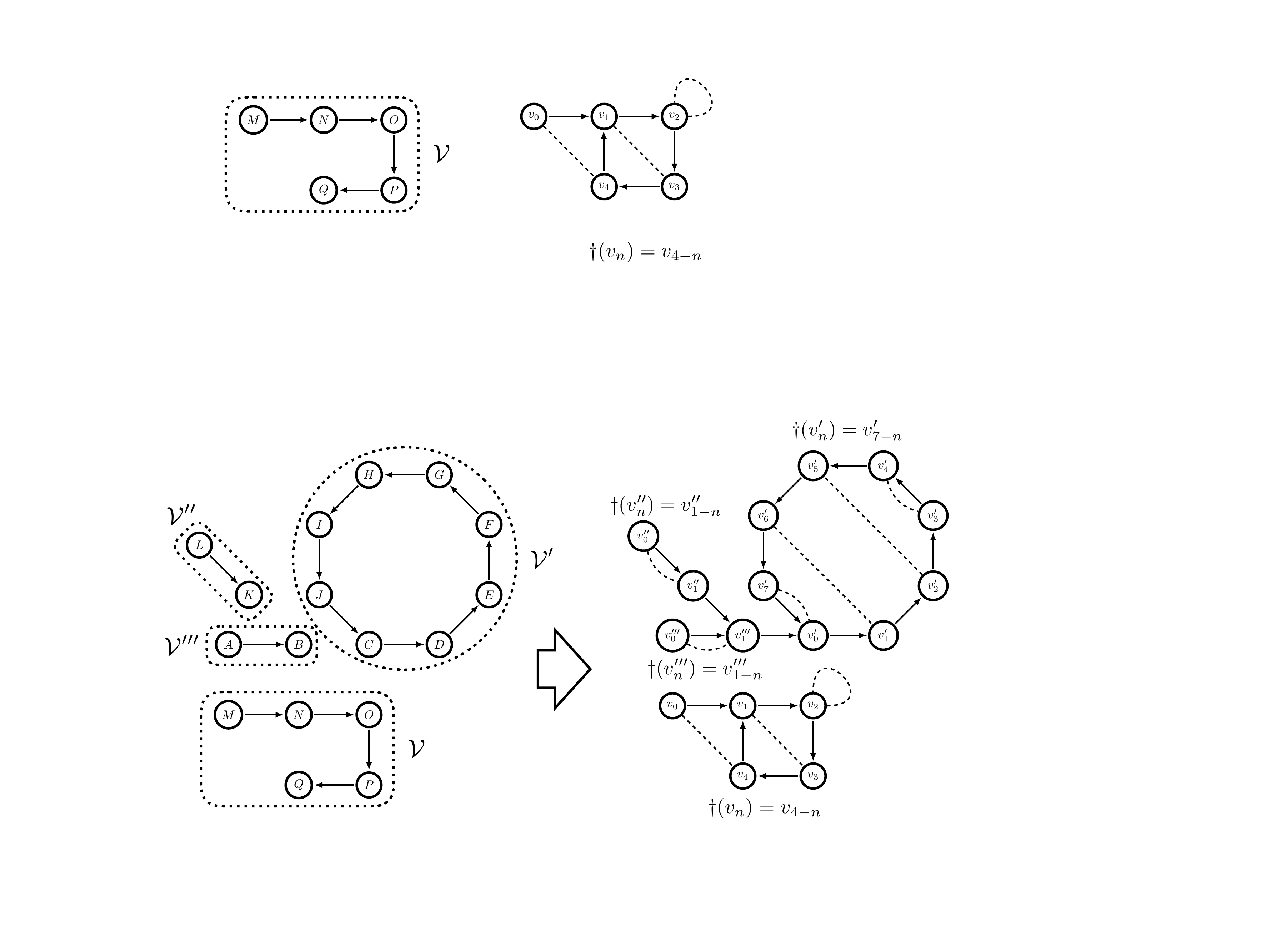}
\caption{For each line or loop component, we relabel the states with $v_n$ such that $C(v_n)=v_{n+1}$.  Then, we define the time reversal involution on each component $\dagger(v_n)=v_{|\mathcal{V}|-n-1}$, shown as the dashed lines between memory states on the right.  As a result, every state, except the last state of a line component $v_{|\mathcal{V}|-1}$, maps to itself via the operation of $\dagger(C(\dagger(C(v_n))))$.  This can be verified for a state in this graph by following a solid directed edge then a dashed bidirectional edge twice.  The solid directed edges on the right diagram represent the memory state transitions of the computation.  Only three states do not return to themselves under the operation $\dagger(C(\dagger(C(v_n))))$, contributing a total of $|\mathcal{M}|-|C(\mathcal{M})|$ unreciprocated transitions.}
\label{fig:EfficientReversal}
\end{figure*}

We use components of the pruned graph to define a time-reversal that minimizes the dissipation.  For a loop or line component $\mathcal{V}$ with $|\mathcal{V}|$ states, we can relabel the memory states with enumeration $\mathcal{V}=\{v_0,v_1,\cdots, v_{|\mathcal{V}|-1}\}$  such that $C(v_n)=v_{n+1}$, except for the last state.   If the component being considered is a loop, then $C(v_{|\mathcal{V}|-1})=v_0$, but if the component is a line then $C(v_{|\mathcal{V}|-1})\neq v_0$.  This can happen either because it transitions to another element of $\mathcal{V}$, or it transitions to a different component $\mathcal{V}'\neq \mathcal{V}$.  It should also be noted that $v_0$ has no antecedent in the original computation if it is part of a line component.

Whether the component $\mathcal{V}$ is a line or a loop, we define the time reversal operation on that component by 
\begin{align}
\dagger (v_n)= v_{|\mathcal{V}|-n-1},
\end{align}
which maps the component to itself.  Note that this operation is an involution $\dagger(\dagger(m))=m$ and can therefore be realized by a memory device described in the previous section.  This transformation can be represented by a bidirectional graph shown by the dashed lines on the right side of Fig.~\ref{fig:EfficientReversal}.  Note, if you follow the series of operations $\dagger(C(\dagger(C(m))))$, the memory state returns to itself, unless it's the final state $v_{|\mathcal{V}|-1}$ in a line component.  

To clarify, there are three cases to consider in evaluating the action of the operator $(\hat{\dagger}\hat{C})^2$:
\begin{enumerate}
\item If $\mathcal{V}$ is either a loop or a line component and the initial memory state $v_n$ is not the last state of the component $v_{|\mathcal{V}|-1}$, then the operation of $(\hat{\dagger}\hat{C})^2$ maps $v_n$ to itself:
\begin{align*}
\dagger(C(\dagger(C(v_n)))) &= \dagger(C(\dagger(v_{n+1}))) 
\\ & =\dagger(C(v_{|\mathcal{V}|-n-2}))
\\ & =\dagger(v_{|\mathcal{V}|-n-1})
\\ & =v_{n}.
\end{align*}

\item If $\mathcal{V}$ is a loop component and the initial memory state is the last state of the component $v_{|\mathcal{V}|-1}$, then the operation of $(\hat{\dagger}\hat{C})^2$ maps $v_{|\mathcal{V}|-1}$ to itself:
\begin{align*}
\dagger(C(\dagger(C(v_{|\mathcal{V}|-1})))) &= \dagger(C(\dagger(v_{0}))) 
\\ & =\dagger(C(v_{|\mathcal{V}|-1}))
\\ & =\dagger(v_0)
\\ & =v_{|\mathcal{V}|-1}.
\end{align*}

\item If $\mathcal{V}$ is a line component and the initial memory state is the last state of the component $v_{|\mathcal{V}|-1}$, then the operation of $(\hat{\dagger}\hat{C})^2$ cannot map $v_{|\mathcal{V}|-1}$ to itself.  If $v_{|\mathcal{V}|-1}$ did map to itself,   it would imply that
\begin{align*}
\dagger(C(\dagger(C(v_{|\mathcal{V}|-1})))) & =v_{|\mathcal{V}|-1},
\end{align*}
which would also imply, by applying $\dagger$ to both sides, that
\begin{align*}
C(\dagger(C(v_{|\mathcal{V}|-1}))) & =v_{0}.
\end{align*}
However, as previously established, the first state $v_0$ in a line component has no antecedents in the computation, so this is impossible.
\end{enumerate}
These results can be confirmed within the example shown in  Fig.~\ref{fig:EfficientReversal} by following an alternating sequence of directed and undirected edges within the diagram, as described in the caption.

Thus, if we calculate the trace in the orthonormal memory-state basis,
\begin{align*}
  \text{Tr}((\hat{\dagger}\hat{C})^2) &=\sum_{m} \bra{m} (\hat{\dagger}\hat{C})^2\ket{m}
  =\sum_{m} \braket{m | (\dagger C)^2(m)} 
  \\ &=\sum_{m} \delta_{m,(\dagger C)^2(m)} ~,
\end{align*}
it returns the number of states that map to themselves under the operation $(\hat{\dagger}\hat{C})^2$, which is $|\mathcal{M}|$ minus the number of end states for line components.  Recall that the number of end states for these components is the same as the number of redundant transistions, and thus $|\mathcal{M}|$ minus the dimension of the image of the computation $|C(\mathcal{M})|$.  Therefore, we have identified an involution $\dagger$ for which $ \text{Tr}((\hat{\dagger}\hat{C})^2)= |C(\mathcal{M})|$.

The combinatoric result that an involution can be found such that $ \text{Tr}((\hat{\dagger}\hat{C})^2)= |C(\mathcal{M})|$ for any $C$ is intriguing on its own.  In the special case that the computation $C$ is a permutation, this means that the operator $(\hat{\dagger}\hat{C})^2$ maps every state to itself, which also means that $\hat{\dagger}\hat{C}$ is itself an involution operator.  This is another proof of the well-known result that any permutation $\hat{C}$ can be expressed as a product of two involutions \cite{pete12a}, in this case $\hat{\dagger}$ and $\hat{\dagger} \hat{C}$.  

However, the physical relevance comes from the results of Sec.~\ref{sec:DesigningInvols}, where we established that any involution is the time reversal of an appropriately designed memory.  We can therefore construct a physically realizable memory device to conduct our computation $C$ while minimizing dissipation divergence according to Eq.~\ref{eq:DissipationBound}.  Thus, the dissipation minimum derived for time-symmetrically controlled operations \cite{Riec19a} obeys the relation
\begin{align}
\lim_{\epsilon \rightarrow 0}\frac{\beta \langle W_\text{diss}\rangle_\text{min} }{\ln (\epsilon^{-1})}  =\frac{|\mathcal{M}|-|C(\mathcal{M})|}{|\mathcal{M}|}.
\end{align}  
The cost of memory state compression hinted at by the device independent inequality in Eq. \ref{eq:DissipationBound} is potentially realizable with a memory designed as described in this chapter.

\section{Conclusion}

Time-symmetric control and metastability apply to a wide range of computations, including biochemical processes and modern digital computers.  As such, the thermodynamic limits on this form of computation should be considered when addressing practical computation.  We see that for nearly-deterministic computations, with small error $\epsilon$, the most important term in the work dissipated is the coefficient of $\ln \epsilon^{-1}$, which we refer to as the dissipation divergence.  We find a simple expression for this divergence in terms of both the computation and the time-reversal operator.  

We explore the relevance of the time-reversal operator by considering different types of memory devices, composed of either magnetic or positional memory.  The different time-reversal symmetries lead to different dissipation divergences and thus different thermodynamic efficiencies that depend on the type of memory device.  We explore some examples, showing that some computations are much more efficiently implemented within magnetic memory devices while others are better suited for positional information storage.  Thus, there is an energetic advantage to using different forms of memory for different computations.

Then, using the flexibility of time-reversal symmetries in memory devices, we ask how much we can minimize dissipation.  We find that the most efficient devices dissipate with divergence proportional to the state compression of the computation.  
Thus, as with Landauer's bound, we see logical reversibility (invertibility) is a necessity to eliminate energy costs of computing.  But, unlike the $k_B T \ln 2$ work required to erase a bit, which can be recovered through quasistatic operations, the logical-irreversibility in time-symmetrically driven computations wastes energy that cannot be recovered.  Moreover, the dissipation diverges with the effectiveness of the computation, elevating the importance of logical reversibility in designing efficient computations.  This suggests that a transition to reversible universal logic gates, like the Fredkin gate \cite{Fred82b, Ray20a}, may provide considerable energetic benefits beyond Landauer's bound.

\section*{Acknowledgments}
\label{sec:acknowledgments}

The authors thank the Telluride Science Research Center for hospitality during visits and the participants of the Information Engines Workshops there for helpful discussions.  We also acknowledge helpful discussions with J. Thompson. This material is based upon work supported by, or in part by, the Templeton World Charity Foundation grants TWCF0560 and TWCF0337, grants FQXi-RFP-IPW-1902 and FQXi-RFP-IPW-1903 from the Foundational Questions Institute and Fetzer Franklin Fund (a donor-advised fund of Silicon Valley Community Foundation), the U.S. Army Research Laboratory and the U. S. Army Research Office under grants W911NF-18-1-0028 and W911NF-21-100048, the National Research Foundation (NRF), Singapore, under its NRFF Fellow program (Award No. NRF-NRFF2016-02), and the Singapore Ministry of Education Tier 1 Grant No. RG146/20.  Any opinions, findings and conclusions or recommendations expressed in this material are those of the author(s) and do not reflect the views of the National Research Foundation, Singapore.

\appendix

\section{Eigenvalues of a Computation}
\label{sec:Eigenvalues of a Computation}

For a computation matrix $\hat{C}$ that is determined by the function on the memory states $C: \mathcal{M} \to \mathcal{M}$, it is possible to separate memory $\mathcal{M}$ into a transient component $\mathcal{M}_\text{T}$ and a recurrent component $\mathcal{M}_\text{R}$.  We define the transient component as the set of states which has zero probability of occupation after $|\mathcal{M}|$ operations $\mathcal{M}_\text{T}=\{m'| \nexists \text{ } m\in \mathcal{M} \text{ such that }  C^{|\mathcal{M}|}(m)=m'\}$.  The choice of the depth $|\mathcal{M}|$ comes from the fact that $|\mathcal{M}|$ is the longest possible chain of states that could be followed through repeated computations without seeing a repetition.  If we made the condition that we operated with the computation fewer times, it would be possible to arrive at a memory state that is never returned to in future iterations.  But all states which satisfy the converse are thus part of the recurrent memory states $\mathcal{M}_\text{R}=\{m'| \exists \text{ } m \in  \mathcal{M} \text{ such that }  C^{|\mathcal{M}|}(m)=m'\}$.  Note that these conditions guarantee that if a memory state is in the recurrent component $m \in \mathcal{M}_\text{R}$, then so is the resulting memory state after computation $C(m) \in \mathcal{M}_\text{R}$.  But, if the memory is in a transient state $m \in \mathcal{M}_\text{T}$, then the computation may lead it out of the set of transient states $C(m) \in \mathcal{M}_\text{R}$.

Let $\hat{I}_\text{R} \equiv \sum_{m \in \mathcal{M}_\text{R}} \ket{m} \bra{m}$ 
be the projection onto the span of the recurrent states $\linspan{\text{R}} = \text{span}(\mathcal{M}_\text{R})$
and let $\hat{I}_\text{T} \equiv \sum_{m \in \mathcal{M}_\text{T}} \ket{m} \bra{m}$ 
be the projection onto the span of the transient states $\linspan{\text{T}} = \text{span}(\mathcal{M}_\text{T})$.
Recall that the net computation operator was defined as
$\hat{C} = \sum_{m \in \mathcal{M}} \ket{C(m)} \bra{m}$.
A proper ordering of memory states allows the computation matrix to be represented as the block matrix:
\begin{align}
\hat{C} = 
\begin{bmatrix}
\hat{C}_\text{T}^{(1)} & \boldsymbol{0} \\
\hat{C}_\text{T}^{(2)} & \hat{C}_\text{R}
\end{bmatrix} ~,
\end{align}
where we define the restricted linear maps
\begin{itemize}
  \item $\hat{C}_\text{R}: \linspan{\text{R}} \rightarrow \linspan{\text{R}}$ with $\ket{m} \mapsto \ket{C(m)}$, \\
  \item $\hat{C}_\text{T}^{(1)}: \linspan{\text{T}} \rightarrow \linspan{\text{T}}$ with $\ket{m} \mapsto \ket{C(m)} \LIverson C(m) \in \linspan{\text{T}} \RIverson$, and \\
  \item $\hat{C}_\text{T}^{(2)}: \linspan{\text{T}} \rightarrow \linspan{\text{R}}$ with $\ket{m} \mapsto \ket{C(m)} \LIverson C(m) \in \linspan{\text{R}} \RIverson$.
\end{itemize}

Ref.~\cite{Riec18a} pointed out that---due to the triangular block-diagonal structure of transition matrices---the eigenvalues of a transition matrix are the union of the recurrent eigenvalues
and transient eigenvalues.
I.e., the eigenvalues $\lambda \in \Lambda_{\hat{C}}$ of $\hat{C}$ satisfy
$| \hat{C} - \lambda \hat{I}| = | \hat{C}_\text{T}^{(1)} - \lambda \hat{I}_\text{T} | \cdot | \hat{C}_\text{R} - \lambda \hat{I}_\text{R} | = 0$,
which implies that
\begin{align}
\Lambda_{\hat{C}} = \Lambda_{\hat{C}_\text{T}^{(1)}} \cup \Lambda_{\hat{C}_\text{R}} ~.
\end{align}

Since $\hat{C}_\text{T}^{(1)}$ is by definition nilpotent, it can only have zero-valued eigenvalues.

$\hat{C}_\text{R}$ is an orthogonal operator since $\hat{C}_\text{R}^\top \hat{C}_\text{R} = \hat{I}_\text{R}$.
Therefore, the eigenvalues of $\hat{C}_\text{R}$ are all roots of unity
(i.e., $\lambda = e^{i 2 \pi / n}$ for $n \in \mathbb{N}$, such that $|\lambda| = 1$).

The takeaway is that all deterministic computations have eigenvalues which are either zero, or roots of unity.  The roots of unity correspond to the eigenvalues of the recurrent permutation at the core of the computation.

\section{Construction of Arbitrary Involutions}
\label{app:Construction of Arbitrary Involutions}

A memory device is characterized by two parts.  First is a physical system, which is a collection of time-even $q_i \in \mathcal{Q}_i$ and time-odd variables $p_j\in \mathcal{P}_j$ that compose to make a system state $s \in \mathcal{S}=\prod_{i} \mathcal{Q}_i \times \prod_{j} \mathcal{P}_j$.  Second is a coarse graining, which is a set of memory states $\mathcal{M}=\{m\}$ that partition the physical space.  This partitioning can be described by a surjective function $\mathfrak{m}:\mathcal{S} \to \mathcal{M}$ where $\mathfrak{m}(s)=\{ s' |  s \in m \text{ and } s' \in m \text{ and }  m \in \mathcal{M} \}$.
Time-reversal, assumed to be defined on system states, acts on a memory state by commuting with the coarse-graining: $\dagger(\mathfrak{m}(s)) = \mathfrak{m}(\dagger(s))$ for each system state $s \in \mathcal{S}$.
Our task here is to find both a physical system and coarse-graining such that time-reversal $\dagger: \mathcal{S} \to \mathcal{S}$ acts on the memory states according to whatever involution $\invol: \mathcal{M} \to \mathcal{M}$ that we desire: $\dagger(\mathfrak{m}(s)) = \invol(\mathfrak{m}(s))$.  In short, we will show that any involution can be realized as the time reversal of a particular memory system.

The first step in our task is to characterize different types of involutions. An arbitrary involution on memory states is composed of two types of operations.
For each memory state, the involution either maps the state to itself, or swaps it with a partner state, mapping it to a different memory state which is in turn mapped to the first. Memory states that are mapped to themselves under the involution are called unchanging states while those that swap with a partner are called swapping states.  
We define any involution which has $M$ unchanging states and $2N$ swapping states as type $(M, N)$.  
This naturally operates on a total $M + 2N$ memory states, where we have only even numbers of swap states because each of these states has a partner.  
Any physical system which supports an involution of type $(M, N)$ also supports any other involution 
with the same number of swapping states and unchanging states.

This can be seen by considering two different involutions $\invol_1$ and $\invol_2$ that are of the same type $(M, N)$ acting on memory states $\mathcal{M}_1$ and $\mathcal{M}_2$, respectively.
We can then define a relabeling bijection between the memory states $r: \mathcal{M}_1 \to \mathcal{M}_2$ that transforms $\invol_1$ into $\invol_2$ in the following way.
For each pair of swapping states $\{x, y\}$ in $\mathcal{M}_2$, choose a unique pair of swapping states $\{a, b\}$ in $\mathcal{M}_1$, and define $r(a) = x$ and $r(b)=y$.
The resulting involution over $\mathcal{M}_2$ can be expressed in terms of the involution over $\mathcal{M}_1$ and the bijection $r$:
\begin{align*}
  \invol_2(x) &= y
  = r(b)
  = r \invol_1(a) \\
  &= r \invol_1 r^{-1}(x)
  ~.
\end{align*}
The same is true for each unchanging state $z$: choose a unique unchanging state $c$ under $\invol_1$ and let $r(c) = z$, resulting in
\begin{align*}
  \invol_2(z) &= z
  = r(c)
  = r \invol_1(c) \\
  &= r \invol_1 r^{-1}(z)
  ~.
\end{align*}
Thus, the entire involution can be expressed $\invol_2 = r \invol_1 r^{-1}$.

Suppose further that we already found a physical system and coarse-graining with function $\mathfrak{m}_1$ such that
time-reversal $\dagger_1$ on the memory states is $\invol_1$,
but that we need to find such a memory device for $\invol_2$.
Then let the second physical system be the same as the first and define the second coarse-graining function $\mathfrak{m}_2$ by simply composing the memory state relabeling $r$ with $\mathfrak{m}_1$: $\mathfrak{m}_2 = r \mathfrak{m}_1$.
We then have for each microstate $s \in \mathcal{S}$:
\begin{align*}
  \dagger\mathfrak{m}_2(s)
  &= \mathfrak{m}_2\dagger(s) \\
  &= r\mathfrak{m}_1\dagger(s) \\
  &= r\dagger\mathfrak{m}_1(s) \\
  &= r\invol_1\mathfrak{m}_1(s) \\
  &= r\invol_1 r^{-1}\mathfrak{m}_2(s) \\
  &= \invol_2\mathfrak{m}_2(s)
  ~,
\end{align*}
so that $\dagger\mathfrak{m}_2 = \invol_2\mathfrak{m}_2$.
In other words, time reversal acts on the memory states of the second memory device in accordance with the second involution.

So to find a memory device where time-reversal generates an arbitrary involution of type $(M, N)$, we need only show how to construct the physical device and coarse-graining that will generate one such involution under time-reversal.

We accomplish this in four stages.
First, we define the physical system and a trivial coarse-graining that has an excessive number of memory states, all of which are all swap states.
Second, we further coarsen the coarse-graining to define the correct number $M$ of unchanging memory states.
Third, we further coarsen the coarse-graining so that either $2N$ swap states are left, for $N\geq 1$, or two are, for $N=0$.
For the case of $N=0$, we apply a fourth procedure to reduce the number of swap states to zero.

As a physical system, we use $S$ magnetic dipoles, where $S$ is any number such that $2^{S-1} \geq M + N$.
For example, we could choose $S = \lceil\log_2(M+N)\rceil + 1$.
This will guarantee we have enough dipoles to construct our ultimate coarse-graining.
Each dipole is restricted to stably exist in only one of two configurations, meaning that we can treat the number of system states for our physical system as $2^S$.
We define our initial coarse-graining to be a trivially fine coarse-graining mapping each of these $2^S$ system states to a unique memory state.
Under time-reversal, each dipole then flips orientation so that every microstate swaps with one other microstate, meaning that there are no unchanging memory states.
So time-reversal acts on our initial memory device as a type $(0, 2^{S-1})$ involution.

Merging memory states $m$ and $m'$ means creating a new memory state $m''=\text{merge}(m,m')=\{s|\mathfrak{m}(s) = m \text{ or } \mathfrak{m}(s) = m'\}$.
The time-reversal of a merged state is the merger of the time-reversals of the original memory states:
\begin{align}
\dagger(\text{merge}(m,m'))&=\{\dagger(s) |\mathfrak{m}(s)= m \text{ or } \mathfrak{m}(s) = m'\}
\\ &=\{s |\mathfrak{m}(s)= \dagger(m) \text{ or } \mathfrak{m}(s) = \dagger(m')\}
\\ & =\text{merge}(\dagger(m), \dagger(m')).
\end{align} 
Thus, if we also merge the time reversal of $m$ and $m'$ into a new memory state $m'''=\dagger(\text{merge}(m,m'))$, then $m''$ and $m'''$ will be each other's time reversal: $m''=\dagger(m''')$ and $m'''=\dagger(m'')$.

In order to get $N$ unchanging memory states, we enact the second stage by merging memory states to define a new coarse-graining.
Consider a pair of memory states $m\neq m'$ that map onto each other under time reversal.
Then the merger $m''$ of these two states maps to itself: $\dagger(m'')=m''$.
By then adding a new memory state $m''$ to $\mathcal{M}$, changing the coarse-graining so that $\mathfrak{m}(s) = m''$ for all $s$ that previously mapped to either $m$ or $m'$, and throwing out the states $m$ and $m'$ from $\mathcal{M}$ (which will no longer have support), we create a new coarse-graining.
This ensures that time reversal acts as a type $(1, 2^{S-1}-1)$ involution.
By then repeating this procedure, we can merge in total $2M$ swap states into $M$ unchanging states.
This gives a time-reversal of type $(M, 2^{S-1}-M)$.

While we now have $M$ unchanging states, we need to end up with only $N$ pairs of swap states, and $2^{S-1} - M \geq N$.
So the third stage requires us to merge $2^{S-1}-M-N$ more pairs of memory states.
This time, however, we need for the resultant memory states to not map onto themselves under time-reversal.
This is achievable by merging memory states that do not map onto themselves,
as well as merging their time-reverses.
That is, we merge a pair of states $m$ and $m'$, where $\dagger m \neq m'$, into a state $m''$,
as well as $\dagger m$ and $\dagger m'$ into $m'''$.
Then $m''\neq m'''$, but $\dagger m'' = m'''$.
We then throw out the original states $m$, $m'$, $\dagger m$, and $\dagger m'$, add the new states $m''$ and $m'''$ to $\mathcal{M}$ and update our coarse-graining to properly map to $m''$ and $m'''$, like before.
By repeating all of this $2^{S-1}-M-N$ times, time-reversal becomes a type $(M, N)$ involution on the memory states for our new memory device.

However, if $N=0$, then we cannot perform the final merge of the third stage because only one pair of swap memory states will be available, leaving us with a time-reversal involution of type $(M, 1)$.
To finish our coarse-graining, we instead need to make two more steps at this point.
We first perform one step of the second stage procedure, leaving us a with time-reversal involution of type $(M+1, 0)$.
Then we merge two of the remaining, unchanging states.
After adding this final merged state, removing the two unchanged states that were merged, and updating our coarse-graining function, we are finally left with a time-reversal involution of type $(M, 0) = (M, N)$.  And so, we have proved that we can construct a memory with any involution as its time-reversal.

\bibliography{chaos,ref}

\end{document}